\documentclass{emulateapj}
\usepackage{apjfonts}
\usepackage{epsfig}
\usepackage{rotating}
\begin{document}

\submitted{The Astrophysical Journal, 625, in press}

\title{
Ultraviolet, X-ray, and Optical
 Radiation from the Geminga Pulsar\footnotemark[1]
\footnotetext[1]{Based on observations made with the NASA/ESA Hubble
                 Space Telescope, obtained at the
                 Space Telescope Science Institute, which is operated by
                 the Association of Universities for Research in Astronomy,
                 Inc., under NASA contract NAS 5-26555. These
                 observations are associated with programs GO-9182
        and GO-9756.}}

\author{O.\ Y.\ Kargaltsev and
G.\ G.\ Pavlov} \email{green@astro.psu.edu, pavlov@astro.psu.edu}
 \affil{Dept.\ of Astronomy and Astrophysics, The
Pennsylvania State University, 525 Davey Lab., University Park, PA
16802}

\author{V.\ E.\ Zavlin}
\email{vyacheslav.zavlin@msfc.nasa.gov}
 \affil{Space Science Dept.,
NASA Marshall Space Flight Center, SD50, Huntsville, AL 35812}

\and

\author{R.\ W.\
Romani} \email{rwr@astro.stanford.edu} \affil{Stanford University,
Dept.\ of Physics, Stanford, CA 94305}

\begin{abstract}
We observed the gamma-ray pulsar Geminga with the
FUV-MAMA and NUV-MAMA detectors of the Space Telescope Imaging
Spectrometer
to measure the Geminga's spectrum  and pulsations in the
ultraviolet.
The slope of the far-ultraviolet (FUV) spectrum is
close to that of a Rayleigh-Jeans spectrum, suggesting that the FUV
radiation is dominated by thermal emission from the neutron star (NS)
surface. The measured FUV flux, $F_{\rm FUV}=(3.7\pm 0.2)\times 10^{-15}$
ergs cm$^{-2}$ s$^{-1}$ in 1155--1702\,\AA\ band,
corresponds to a brightness temperature $T_{\rm RJ}\approx
(0.3$--$0.4)\,(d_{200}/R_{13})^2$  MK, depending on the interstellar
extinction ($d=200\,d_{200}$ pc and $R = 13\,R_{13}$ km are the
distance and the NS radius).
The soft thermal component of the Geminga's X-ray spectrum measured
with the {\sl XMM-Newton} observatory corresponds to a temperature
$T_s = 0.49\pm 0.01$ MK and radius $R_s = (12.9\pm 1.0)\,d_{200}$
km. Contrary to other NSs detected in the UV-optical, for which
the extrapolation of X-ray thermal component into the optical
underpredicts the observed flux of thermal radiation, the FUV
spectrum of Geminga lies slightly below the extrapolation of the
soft thermal component, which might be associated with the Geminga's
very low temperature. Surprisingly, the thermal FUV radiation is
strongly pulsed, showing a narrow dip at a phase close to that of
a broader minimum of the soft X-ray light curve.
The strong
pulsations might be attributed to
partial occultations of the thermal UV radiation by regions of the
magnetosphere filled with electron/positron plasma.

In contrast with the FUV spectrum,
the near-infrared (NIR) through near-ultraviolet (NUV) spectrum of Geminga
is clearly nonthermal. It
can be described by a power-law model,
$F_\nu \propto \nu^{-\Gamma+1}$,
with a
photon index $\Gamma=1.43\pm 0.15$,
close to the slope
$\Gamma = 1.56\pm 0.24$ of the hard X-ray
($E>2.5$ keV) magnetospheric component. The extrapolation of the
X-ray magnetospheric spectrum into the optical is marginally
consistent with (perhaps lies slightly above) the observed NIR-optical-NUV
spectrum. The NUV pulsations, however, do not show a clear correlation
with the hard X-ray pulsations.
\end{abstract}

\section{Introduction}

Spin-powered pulsars show highly pulsed emission from the radio
to $\gamma$-rays, arising from acceleration zones in their
magnetospheres. In the UV to soft X-ray band, however, thermal emission
from the
neutron star (NS) surface can contribute significantly for middle-aged pulsars,
with characteristic ages $\tau\sim 10^4$---$10^6$ yr.
Spectral and timing measurements can
separate
these
two components, allowing a measure of the surface temperature and thermal
luminosity.
By measuring thermal emission as a function of age, one can probe the equation
of state of matter at supranuclear densities in the NS core and
constrain the surface composition.
Observations with the {\sl Chandra} and {\sl XMM-Newton} X-ray observatories
have begun to reveal much about the thermal
component (see Pavlov, Zavlin, \& Sanwal 2002 and Kaspi, Roberts, \& Harding 2004
for recent reviews).
However, since typical effective temperatures of middle-aged pulsars
are as low as
$\sim 30$--100 eV, and interstellar absorption severely attenuates the flux
below $\sim 0.1$ keV, the X-ray observations of these objects
can only probe
the Wein tail of the surface thermal spectrum. Two issues then complicate the
interpretation. First, surface composition can dramatically affect the X-ray
flux (Romani 1987; Zavlin \& Pavlov 2002)
with light element
atmosphere leading
to a large Wien excess. Second, any surface
temperature inhomogeneities will also complicate the spectrum, with hot spots
disproportionally important in the high energy (X-ray) tail.
        For these reasons, comparison of the X-ray results with UV
emission from the Rayleigh-Jeans side of the thermal bump is particularly
valuable. The challenge here is that non-thermal magnetospheric emission
becomes increasingly dominant as one moves to the red
(Pavlov et al.\ 2002). Fortunately, the
NUV-MAMA and FUV-MAMA detectors of the Space Telescope Imaging Spectrometer
(STIS) aboard the {\sl Hubble Space Telescope} ({\sl HST})
offer access to the UV emission and
provide the
phase-resolved
measurements that
can help to separate the thermal and non-thermal fluxes.

We report here on {\sl HST} STIS observations of the
middle-aged $\gamma$-ray pulsar Geminga.
Discovered in 1972 by the {\sl SAS}-2 satellite
(Fichtel et al.\ 1975), this object had been known only as a $\gamma$-ray
source until it was detected in X-rays by the {\sl Einstein} observatory
(Bignami, Caraveo, \& Lamb 1983) and associated with a faint
($V\approx 25.5$) optical counterpart
(Bignami et al.\ 1987;
Bignami, Caraveo, \& Paul 1988; Halpern \& Tytler 1988).
The discovery of
a period $P=237$ ms in X-rays with {\sl ROSAT} (Halpern \& Holt 1992)
and $\gamma$-rays with {\sl Compton Gamma-ray Observatory} ({\sl CSGO})
(Bertch et al.\ 1992) proved the source to be a spin-powered pulsar,
with a characteristic age $\tau\equiv P/(2\dot{P})=342$ kyr and
 spin-down energy loss rate $\dot{E}=3.3\times 10^{34}$ erg s$^{-1}$.
Contrary to most spin-powered pulsars, Geminga is not a strong
radio source. Detection of pulsed radio emission at 102 MHz was claimed
by Malofeev \& Malov (1997), Kuz'min \& Losovskii (1997), and Shitov
\& Pugachev (1997), but the pulsar has not been detected at other
frequencies (e.g., McLaughlin et al.\ 1999).

{\sl ROSAT}, {\sl EUVE}, and {\sl ASCA} observations have established
that the X-ray spectrum of Geminga consists of a  soft thermal component,
likely emitted from the NS surface, and a nonthermal component,
presumably generated in the pulsar magnetosphere
(Halpern \& Ruderman 1993; Halpern \& Wang
1997; Jackson et al.\ 2002). Recent observation of Geminga with
{\sl XMM-Newton} have shown an extended emission resembling a
bow-shock nebula (Caraveo et al.\ 2003). From a two-component,
blackbody (BB) plus power-law (PL), fit of the phase-integrated
{\sl XMM-Newton} spectrum, Zavlin \& Pavlov (2004a) found a
temperature $T_{\rm bb}\approx 0.5$ MK for the thermal component
and a photon index $\Gamma\approx
2$ for the magnetospheric
component. The X-ray pulse profile shows a strong dependence on
energy, changing from a single broad peak
at $E\lesssim 0.8$ keV to a double-peak
structure
at $E\gtrsim 2$ keV.

The shape of the Geminga's optical spectrum remains controversial.
Based on photometry with a few broad-band filters,
Bignami et al.\ (1996) proposed a broad emission feature around
$\sim 5000$ \AA, superimposed on a Rayleigh-Jeans thermal
spectrum (see also Mignani, Caraveo, \& Bignami 1998),
and interpreted the feature as an ion cyclotron line emitted from the NS
atmosphere. Martin, Halpern, \& Schiminovich (1998) reported
a possible broad dip over 6300--6500 \AA\ in a flat
($\Gamma\approx 1.8$) spectrum spanning 3700--8000 \AA, but the spectrum
was severely contaminated by the sky background.
Harlow, Pavlov, \& Halpern (1998) detected Geminga in two near-IR
bands, which proved that the spectrum grows toward lower
frequencies, similar to another middle-aged pulsar
B0656+14 (Koptsevich et al.\ 2001). Overall, it is clear that the
optical spectrum is predominantly nonthermal, perhaps with a hint
of a Rayleigh-Jeans component at $\lambda\lesssim 3000$ \AA.
Optical pulsations of Geminga were (marginally) detected in the B band only
(Shearer et al.\ 1998).

Based on
three
{\sl HST} WFPC2 observations, Caraveo et al.\ (1996) found
Geminga's parallax of $6.4\pm1.7$ mas, corresponding to $d\approx 160$ pc.
Our reanalysis of these data together with fourth WFPC2 observation
shows that the result is not reliable because the exposures
were too short to determine the Geminga's positions with the required
accuracy (Pavlov et al.\ 2005, in preparation). Therefore, the
distance to Geminga is currently unknown.
In this paper we will scale the distance to $d=200$ pc.

Particularly interesting would be an observation of Geminga
in the far-ultraviolet (FUV) range
where one could expect thermal radiation from the NS surface to
take over the apparently nonthermal radiation that prevails
in the optical. Moreover,
observing
pulsations of Geminga
shortward of $\sim 4000$ \AA\
allow one to study the transformation
of the pulse profile in the transition from the nonthermal to thermal
regime and elucidate the nature of the Geminga's radiation in the
optical-UV range. To measure the spectrum and pulsations in the
ultraviolet, we carried out an imaging observation with NUV-MAMA
and a spectral observation with FUV-MAMA, both with time resolution
of 125\,$\mu$s.
These observations and the data analysis, including
the NIR-optical-UV data, are described in \S2.
In \S3 we present the spectral and timing analyses of a recent
{\sl XMM-Newton} observation of Geminga. Implications of the
broad-band (NIR through X-rays) observations are discussed in \S4.
The results of our work are summarized in \S5.
\section{Observations and Data Analysis}

\subsection{NUV-MAMA photometry}

Geminga was observed with the STIS NUV-MAMA on 2002 February 27 (start date is
$52,332.4340$ MJD UT).
The broad-band filter F25SRF2 (pivot wavelength 2299 \AA, FWHM 1128 \AA)
was used in this imaging observation
to minimize the contribution of geocoronal lines.
The data were taken during four consecutive orbits.
The total scientific exposure time was 11,367 s.

\begin{figure}[]
 \centering
\includegraphics[width=3.3in,angle=0]{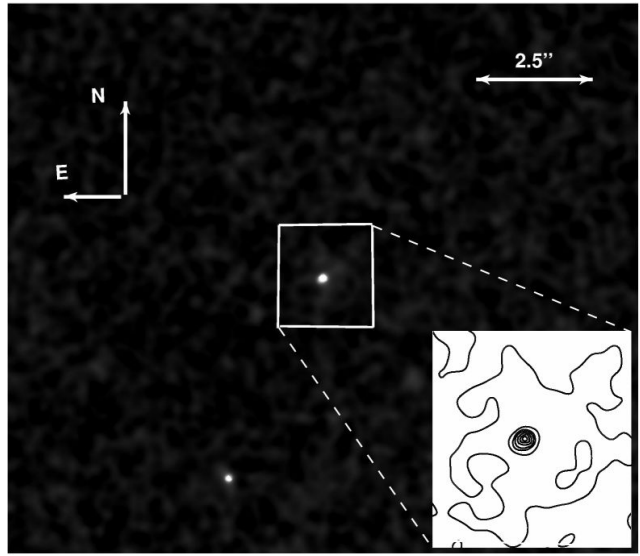}
\figcaption{NUV-MAMA image of the field around the Geminga pulsar
(at the center of the image). The only other point source in the
field is Star G (e.g., Halpern, Grindlay, \& Tytler 1985), used for
acquisition. The inset shows brightness contours in the
$2\farcs1\times 2\farcs2$ region centered on Geminga.}
\end{figure}

To avoid possible additional errors associated with the pipeline
subtraction of the strong dark current background (see \S\,7.4.2 of
the STIS Instrument Handbook [IHB]\footnote[1]{STIS IHB available at
http://www.stsci.edu/hst/stis/documents/handbooks/\\currentIHB/stis\_ihbTOC.html}),
we reprocessed the ``raw" NUV-MAMA images repeating all standard
calibration pipeline steps except for this subtraction. As an
output, we obtained four flat-fielded ``low-resolution'' images
($1024\times1024$ pixels; plate scale $0\farcs0244$ pixel$^{-1}$).
The target was detected in each of the four exposures.
  To increase the signal-to-noise ratio, S/N, we combined the
images from the four exposures into a single image using the
STSDAS\footnote[2]{Space Telescope Data Analysis System available at
\\ http://www.stsci.edu/resources/software\_hardware/stsdas}
 task {\it mscombine}. From the sharpness of the source counts
 distributions for the two point sources detected,  Geminga
 and star G
  (see Fig.\ 1) we conclude
 that the images are aligned well enough
 for the photometry purposes.
(Slight apparent elongations of the images of the two point sources,
in different directions for Geminga and star G, are likely
caused by nonuniformities of the background.)
We measure the background, which is dominated by the detector dark
current,
in the annulus with the
 inner radius of 40 pixels and outer radius of 55 pixels, centered
  on the source ($X=539.5$ pixels, $Y=540.5$ pixels).
The mean background count rate within the annulus is
$1.59\times 10^{-3}$ counts s$^{-1}$ pixel$^{-1}$.

To find an optimal aperture radius,
we measured the number of source counts,
$N_s=N_t - N_b$
(where $N_t$ is the total number of counts, and $N_b$ is the number
of background counts estimated by scaling the mean background in the
annulus to the aperture area), and its uncertainty, $\delta N_s$,
 in apertures with radii of about 2, 3, 4, 5, 6, and 7 pixels
(see Table 1).
 To evaluate the background uncertainty needed for calculating $\delta N_s$,
we put each of the apertures at 15 positions randomly distributed
over the annulus, measured the number of background counts within the
aperture for each position, and calculated the root-mean-square, $\delta N_b$, of
the differences between this number and the mean background scaled to
the aperture area.
The uncertainty of the source counts
can then be calculated as $\delta N_{s} = [N_{s} +
\delta N_b^2 (1+1/15)]^{1/2}$ for each of the apertures. From
Table 1 we see that the dependence of S/N ($= N_s/\delta N_s$) on
aperture radius has a flat maximum at a level of ${\rm S/N}\approx
35$ at $r\approx 3$--6 pixels.

\begin{table}[tbp]
\caption[]{Numbers of NUV counts for different extraction aperture
sizes}
\begin{center}
\begin{tabular}{ccccccccccc}
\tableline\tableline $r_{s}$\tablenotemark{a}  & $N_{t}$ &
$N_{b}$\tablenotemark{b} & $\delta N_{b}$   & $N_{s}$ & $\delta
N_{s}$ & S/N & $\bar{\epsilon}$& $C$\tablenotemark{c} &
$\bar{F}_\lambda$\tablenotemark{d}
& $\langle F_\lambda\rangle$\tablenotemark{d}\\
\tableline
      1.95 &         1464.7 &      215.4 &      40.2 &      1249.2
&      54.5 &      22.9&  0.496 &0.222   & 1.26 & 1.33 \\
      2.99 &         2119.6 &      502.7 &      21.8 &      1616.9
&      46.1 &      35.1& 0.566 & 0.251& 1.45 & 1.50 \\
      3.95 &         2688.3 &      879.7 &      26.1 &      1808.6
&      50.4 &      35.9& 0.628 &0.253& 1.47 & 1.51 \\
      4.98 &         3357.4 &      1400.3 &      35.2 &      1957.1
&      57.3 &      34.2& 0.691 &0.249& 1.46 & 1.49 \\
      5.97 &         4115.1 &      2010.6 &      35.4 &      2104.5
&      58.7 &      35.9& 0.716 &0.259& 1.51 & 1.55 \\
      7.02 &          4950.4 &      2782.6 &      51.2 &      2167.8
&      70.5 &     30.8& 0.741 &0.257& 1.50 & 1.53 \\
     \tableline
\end{tabular}
\end{center}
\tablenotetext{a}{Radius of the extraction aperture in pixels.}
\tablenotetext{b}{Number of background counts within the extraction
aperture.} \tablenotetext{c}{Source count rate corrected for the
finite aperture in counts s$^{-1}$.} \tablenotetext{d}{Mean spectral
flux (see eqs.\ [2] and [3]) in units of $10^{-18}$ ergs cm$^{-2}$
s$^{-1}$ \AA$^{-1}$.}
\end{table}

We also measured the numbers of counts in the image combined from
the automatically processed images (with the dark current
subtracted), performing standard aperture photometry with the IRAF
task {\it phot} from the {\it apphot} package\footnote[3]{
http://stsdas.stsci.edu/cgi-bin/gethelp.cgi?phot.hlp}. A good
agreement with the results obtained from the direct measurements of
the total (dark current plus sky) background (e.g., $N_s\pm \delta
N_s = 1821\pm 51$ vs.\ $1808\pm 50$, for the 4-pixel-radius
aperture) shows that the pipeline subtraction of the dark current
does not introduce substantial errors in this case.

The source spectral flux $F_\lambda$ is connected with the number of source
counts in a given aperture by the integral relation
\begin{equation}
N_s = t \int R_\lambda \lambda F_\lambda \epsilon_\lambda\,\,d\lambda~,
\end{equation}
where $t$ is the exposure time,
 $R_\lambda$ is the integrated system throughput, including
the Optical Telescope Assembly (OTA)
 and filter throughputs\footnote[4]{We corrected
the throughputs supplied with the data for the time-dependent
sensitivity loss (see
http://www.stsci.edu/hst/stis/calibration/\\reference\_files/tds.html).},
and $\epsilon_\lambda$ is the wavelength-dependent encircled energy
fraction. One can define the average flux in the filter passband as
either
\begin{equation}
\bar{F}_\lambda = \frac{N_s}{t\int R_\lambda \lambda \epsilon_\lambda\,\, d\lambda}
\end{equation}
or
\begin{equation}
\langle F_\lambda \rangle =
\frac{N_s}{t\bar{\epsilon}\int R_\lambda \lambda\,\, d\lambda}\,,
\end{equation}
where $\bar{\epsilon}$ is in average encircled energy fraction in the
filter passband, and $N_s/(t\bar{\epsilon}) = C$ is the source count rate
corrected for the finite aperture.
We calculated the average spectral fluxes in both ways (see Table 1)
using the $\epsilon_\lambda$ values
measured by Proffitt et
al.\ (2003) for several aperture radii.
We see that the mean fluxes,
$\bar{F}_\lambda \simeq
\langle F_\lambda\rangle\simeq 1.5\times 10^{-18}$ ergs cm$^{-2}$ s$^{-1}$ \AA$^{-1}$
are close to each other for $r\gtrsim 3$ pixels.
The uncertainty of these values, $\sim 10\%$, is mostly due to
systematic errors in the encircled energy fraction.

\begin{figure}[]
 \centering
\includegraphics[width=2.5in,angle=90]{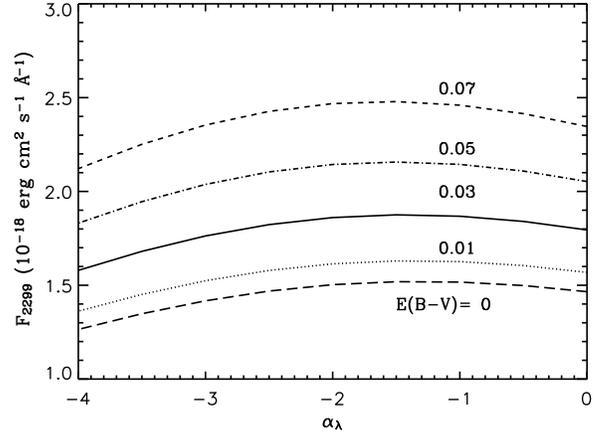}
\figcaption{ Spectral flux $F_\lambda$ at $\lambda=2299$\,\AA\ as a
function of spectral slope $\alpha_\lambda$ in the NUV-MAMA/F25SRF2
band, for different values of E(B-V).}
\end{figure}

Another way to evaluate the flux is to assume some shape for the
spectral flux $F_\lambda$ and determine its normalization making use
of equation (1). We approximate the spectral flux in the F25SRF2
passband as an absorbed power law:
$F_{\lambda}=F_{2299}\,(\lambda/2299{~\rm \AA})^{\alpha_{\lambda}}\,
10^{-0.4 A(\lambda)E(B-V)}$, where $F_{2229}$ is the intrinsic
source spectral flux at the pivot wavelength [it coincides with
$\bar{F}_\lambda$ in the special case $\alpha_{\lambda}=0$,
E(B$-$V)=0], and $A(\lambda)$ is the ultraviolet extinction curve
(Seaton 1979). The color index E(B$-$V) is poorly known. An estimate
based on the hydrogen column density found from the X-ray fits (see
\S3.1) gives E(B$-$V)$\simeq 0.03$; below we will adopt
E(B$-$V)=0.01--0.07 as a plausible range. We calculated the
dependencies of $F_{2299}$ on the spectral slope $\alpha_{\lambda}$
in
 a reasonable range $-4 \leq \alpha_{\lambda} \leq 0$ for several values of E(B$-$V),
based on the $N_s$ values measured in the 4-pixel radius aperture
(see Fig.\ 2).
We see that, at a given E(B$-$V), $F_{2299}$ varies with $\alpha_{\lambda}$ by up to 20\%.
We estimate the uncertainty of the
$F_{2299}$ values at given $\alpha_{\lambda}$ and E(B$-$V)
as $\approx 8\%$--10\%, mostly associated
with
changes of the MAMA imaging point spread function (PSF) between
individual observations that cause systematic uncertainties of
$\epsilon_\lambda$ (see Proffitt et al.\ 2003 and \S16.1 of IHB).

\subsection{FUV-MAMA spectrum}
Geminga was observed with the STIS FUV-MAMA on 2002 February 26 (start date is
$52,331.2391$ MJD UT).
The low-resolution
 grating G140L (which covers the wavelength interval $\approx1150$--1700 \AA)
with the $52'' \times 0\farcs5$ slit was used.
 The data were taken
during four consecutive orbits (including the target acquisition).
We used a nearby field star G (V=21.3; Fig.\ 1) as the acquisition
target and applied a 4\farcs9
offset,
deduced from the positions of
Geminga and the acquisition star measured in the archival {\sl
HST} images, which placed Geminga about 0\farcs1 off the slit
center.
The total scientific exposure time was 10,674 s.

For each exposure, we processed the raw ``high-resolution''
images ($2048\times2048$ pixels; plate scale of $0\farcs0122$ per
pixel
--- see \S11 of the STIS IHB) using the calibration files available on
2003 July 1.
As an output, we obtained flat-fielded low-resolution
($1024\times1024$ pixels; plate scale $0\farcs0244$ pixel$^{-1}$;
spectral resolution 0.58 \AA\ pixel$^{-1}$) images and used them
for the spectral analysis.

The processed images show a nonuniform detector background that
consists of a flat (constant) component and the so-called
``thermal glow'' component (Landsman 1998) that dominates over
most of the detector area and grows with increasing the temperature of
the FUV-MAMA low-voltage power supply (LVPS)
(the average LVPS temperatures were 38.45, 39.67, 40.89, and
41.62 C in the four consecutive orbits of our observation).
 The thermal
glow is the strongest in the upper-left quadrant of the detector,
where the dark count rate can exceed the nominal value,
$6\times10^{-6}$ counts s$^{-1}$ pixel$^{-1}$, by a factor of 20.
To reduce the contamination caused by the thermal glow background,
the source was placed close to the bottom edge of the detector
(see Fig.\ 3).

\begin{figure}[]
 \centering
\includegraphics[width=3.3in,angle=0]{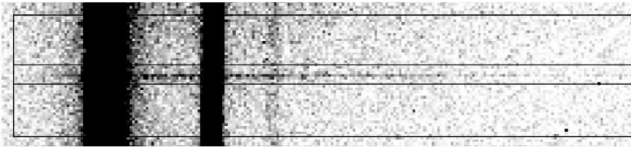}
\vspace{0.0cm} \figcaption{Raw FUV-MAMA spectrum of Geminga. The
boxes show approximate regions for the source and background
extraction used in spectral analysis. }
\end{figure}

We find the Geminga's spectrum centered at $Y=105\pm 2$ pixels in each of
the flat-fielded images (the centroid position slightly varies with $X$),
where $X$ and $Y$ are the image coordinates along the dispersion
and spatial axes, respectively.  Even at this location on the
detector the background still exceeds the nominal value by a
factor of 1.5--5, depending on the position along the dispersion
axis. To improve S/N,
we combined the
images from four exposures into a single image using the STSDAS
 task {\it mscombine}.
The $Y$-positions of the centroids differ by less than 3 pixels
for different exposures and different wavelengths ($X$-positions).

Accurate subtraction of the enhanced, nonuniform background
[typical values are
(1--3) $\times 10^{-5}$ counts s$^{-1}$
pixel$^{-1}$]
is crucial to measuring the spectrum of our faint target. The
spectral extraction algorithm implemented in the standard STIS
pipeline (task X1D) does not adequately correct for the nonuniform
background while extracting the spectrum of such a faint source
and does not allow to vary the extraction box size with the
position along the dispersion axis. Therefore, we
used an IDL
routine with additional capabilities of grouping and fitting the
background and selecting an optimal extraction box size depending
on the position along the dispersion axis
(see Kargaltsev, Pavlov, \& Romani 2004).

Since the source spectrum occupies only a small region on the
detector, we do not attempt to subtract the background globally.
Instead, we scan the count distribution within two strips, $36\leq
Y\leq 95$ and $116\leq Y \leq 175$, adjacent to the source region,
$96\leq Y\leq 115$. To obtain the spectrum with a sufficiently
high S/N, we have to bin the spectrum heavily; after some
experimenting, we chose 12 spectral bins ($\lambda$-bins;
see Table 2).
The bins exclude the regions contaminated by the geocoronal
emission
(Ly${\alpha}$ line and OI lines at 1304 \AA\ and 1356
\AA) and by an artificial background structure at $\lambda\approx 1379$--1384 \AA,
$Y\approx 96$--103. The bins outside the contaminated regions were chosen
to have comparable S/N ($\approx 6$--8), whenever possible.

\begin{table}[tbp]
\caption[]{FUV-MAMA counts and fluxes in $\lambda$-bins}
\begin{center}
\begin{tabular}{ccccccccc}
\tableline\tableline $\lambda$-bin (\AA) & $A_s$\tablenotemark{a} &
$N_{t}$ & $N_{b}$ & $\delta N_{b}$  & $N_{s}$ & $\delta N_{s}$
& S/N & $\langle F_\lambda\rangle \pm \delta \langle F_\lambda\rangle$\tablenotemark{b} \\
\tableline
 1155$-$1187 &  9&    238.3 &      157.3 &     17.2 &     81.1 &   19.9 &  4.1 & $ 13.8 \pm  3.4$ \\
 1248$-$1259 &  9&    130.7 &      63.6 &      5.39 &     67.1 &   9.9 &   6.8 & $ 9.5  \pm  1.4$ \\
 1260$-$1270 &  9&    110.9 &      50.7 &      5.23 &     60.2 &   9.5 &   6.4 & $ 7.6  \pm  1.2$ \\
 1271$-$1283 &  9&    132.9 &      54.0 &      4.97 &     78.9 &   10.3 &  7.7 & $ 9.8  \pm  1.3$ \\
 1316$-$1332 &  7&    130.3 &      57.2 &      8.14 &     73.1 &   12.0 &  6.1 & $ 7.1  \pm  1.2$ \\
 1333$-$1347 &  11&   145.8 &      67.5 &      8.66 &     78.3 &   12.6 &  6.2 & $ 8.2  \pm  1.3$ \\
 1365$-$1378 &  11&   96.6 &      49.3 &      7.11 &     47.3 &   10.1 &  4.7 & $ 5.3  \pm  1.1$ \\
 1385$-$1402 &  11&   123.6 &      62.0 &      4.82 &     61.6 &   9.3 &   6.6 & $ 6.8  \pm  1.0$ \\
 1403$-$1431 &  5&    112.5 &      42.6 &      6.35 &     69.9 &   10.6 &  6.6 & $ 6.3  \pm  1.0$ \\
 1432$-$1471 &  5&    155.8 &      53.1 &      6.98 &     102.8 &   12.4 &  8.3 & $ 7.7  \pm  0.9$ \\
 1472$-$1525 &  5&    140.1 &      61.3 &      8.01 &     78.9 &   12.1 &  6.5 & $ 6.0  \pm  0.9$ \\
 1526$-$1702 &  5&    258.5 &      155.1 &     9.39 &     103.4 &  14.0 &  7.4 & $ 5.1  \pm  0.7$ \\
Summed\tablenotemark{c}& ... & 1776.0& 873.6  & 28.8  & 902.5 & 42.3\tablenotemark{d} & 20.9 & $6.76\pm0.43$\tablenotemark{e} \\

\tableline
\end{tabular}
\end{center}
\tablenotetext{a}{Height of extraction box, in pixels.}
\tablenotetext{b}{Average spectral flux and its statistical error,
in units of $10^{-18}$ erg s$^{-1}$ cm$^{-2}$ \AA$^{-1}$, corrected
for the finite aperture.} \tablenotetext{c}{Values for summed
$\lambda$-bins.} \tablenotetext{d}{Defined as $[\sum_i (\delta
N_{s,i})^2]^{1/2}$.} \tablenotetext{e}{Defined as $\left[\sum_i
\langle F_\lambda\rangle_i \Delta\lambda_i \pm \left(\sum_i \langle
\delta F_\lambda\rangle_i ^{2} \Delta\lambda_i^2\right)^{1/2}\right]
\left(\sum_i \Delta\lambda_i\right)^{-1}$.}

\end{table}

For each of the $\lambda$-bins, we calculate the total number of
counts, $N_t$, within the extraction boxes of different heights
(one-dimensional apertures): $A_{s}=3$, 5, 7, 9, 11, 13, 15, and
17 pixels, centered at $Y=106$ for the first two $\lambda$-bins
and at $Y=105$ for the rest of the $\lambda$-bins. To evaluate the
background, we first clean the background strips (see above) from
outstanding ($>10^{-3}$ cts s$^{-1}$ pixel$^{-1}$) values (``bad
pixels'') by setting them to local average values.
Then, for each of the
$\lambda$-bins, we fit the $Y$-distribution of the background
counts with a first-order polynomial (interpolating across the
source region), estimate the number $N_b$ of background counts
within the source extraction aperture $A_s$, and evaluate the
number of source counts, $N_{s}=N_t-N_b$ (Table 2).

The uncertainty $\delta N_s$ of the source counts can be evaluated
as $\delta N_s = [N_s + \delta N_b^2(1+A_s/A_b)]^{1/2}$, where
$\delta N_b$ is the background uncertainty in the source aperture.
We binned the distribution of background counts along the $Y$-axis
with the bin sizes equal to $A_s$ and calculated $\delta N_b$ as
the root-mean-square of the differences between the actual numbers of
background counts in the bins and those obtained from the fit to
the background.
 We calculated $\delta N_s$ and S/N for various extraction box heights
and found
the $A_s$ values maximizing S/N
for each $\lambda$-bin
(see Table 2).

\begin{figure}[]
 \centering
\includegraphics[width=2.5in,angle=90]{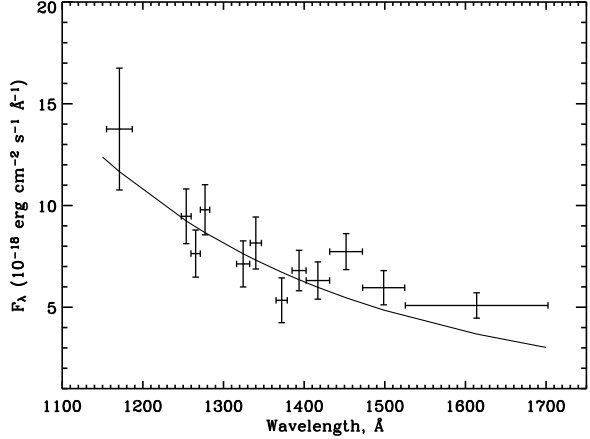}
\caption{ The measured (absorbed) FUV-MAMA spectrum of Geminga. The
solid curve
 shows the best-fit absorbed blackbody model
for E(B$-$V)=0.03 ($T=0.31$ MK at $R=13 d_{200}$ km). }
\end{figure}

We calculated the average spectral fluxes in the $\lambda$-bins
(cf.\ eq.\ [3]):
\begin{equation}
 \langle F_{\lambda}\rangle_i =
\frac{\int_{\Delta\lambda_i} R_{\lambda}\lambda\ F_{\lambda}\,{\rm
d}\lambda} {\int_{\Delta\lambda_i} R_{\lambda}\lambda\, {\rm
d}\lambda} = \frac{C_i}
{\int_{\Delta\lambda_i}
 R_{\lambda}\lambda\, {\rm d}\lambda}\, ,
\end{equation}
where
$C_i$ is the source count rate in the $i$-th
 $\lambda$-bin corrected for the finite size of the source extraction aperture, and
$R_{\lambda}$ is the system response that includes the OTA
throughput and accounts for the grating and slit losses and
time-dependent sensitivity losses (Bohlin, 1999; see also \S 3.4.12
of the HST Data Handbook for STIS\footnote[5]{
http://www.stsci.edu/hst/stis/documents/handbooks/currentDHB/\\STIS\_longdhbTOC.html}
for details). The resulting flux values are given in Table 2, while
the spectrum is shown in Figure 4. The total flux in the 1155--1702
\AA\ range $(\Delta\lambda=547$\,\AA),
 can be estimated as
$F\simeq \Delta\lambda\, \left(\sum_i \langle F_\lambda\rangle_i
\Delta\lambda_i\right) \left(\sum_i \Delta\lambda_i\right)^{-1}
\simeq (3.72\pm 0.24) \times 10^{-15}$ erg s$^{-1}$ cm$^{-2}$,
corresponding to the luminosity $L_{\rm FUV}=4\pi d^2 F =(1.78\pm
0.11)\times 10^{28} d_{200}^2$ erg s$^{-1}$.

\begin{figure}[]
 \centering
\includegraphics[width=2.6in,angle=90]{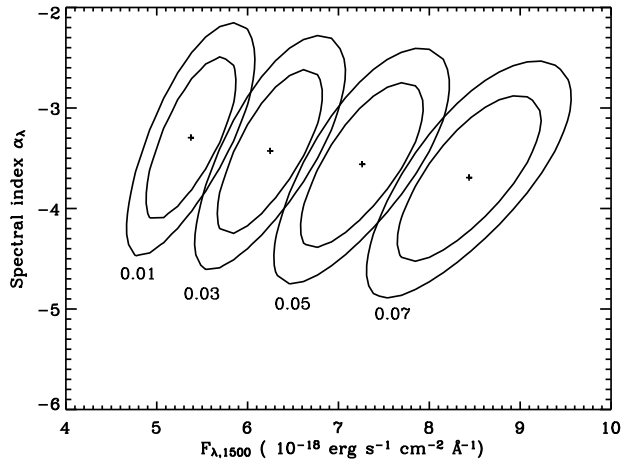}
\caption{ Confidence contours (67\% and 90\%) for the absorbed
power-law model fit to the FUV-MAMA spectrum, for E(B$-$V)=0.01,
0.03, 0.05, and 0.07.}
\end{figure}

We fit the spectrum with the absorbed power-law model,
$F_{\lambda}= F_{1500}\,(\lambda/1500\, {\rm
\AA})^{\alpha_{\lambda}}\times 10^{-0.4 A(\lambda)\, E(B-V)}$.
For plausible values E(B$-$V) = 0.01, 0.03, 0.05, and 0.07, we found
the power-law indices $\alpha_\lambda = -3.29\pm 0.53$, $-3.43\pm
0.53$, $-3.56\pm 0.54$, and $-3.69\pm 0.54$, and the normalizations
$F_{1500} = 5.38\pm0.33$, $6.25\pm0.38$, $7.26\pm 0.37$ and
$8.44\pm 0.52 \times 10^{-18}$ erg cm$^{-2}$ s$^{-1}$ \AA$^{-1}$,
respectively (Fig.\ 5);
the corresponding $\chi_\nu^2$ values are
0.80, 0.81, 0.81, 0.82, and 0.83, for
10 degrees of freedom (dof).

 The inferred slope $\alpha_\lambda$ is close to that of the
Rayleigh-Jeans spectrum, $F_\lambda\propto \lambda^{-4}$,
suggesting that the observed radiation is dominated by thermal emission
from the NS surface.
To estimate the NS surface temperature,
we fit the absorbed blackbody model to the
observed spectrum.
Since the FUV fluxes are in the Rayleigh-Jeans part of the spectrum,
the temperature is strongly correlated with the radius-to-distance
ratio (approximately, $T\propto d^2/R^2$), as demonstrated by the confidence
contours in the $T$-$R$ plane
(Fig.\ 6).
For a typical
 NS radius $R=13$ km and the assumed distance $d=200$ pc,
the inferred surface temperatures are
$0.27\pm 0.01$, $0.31\pm
0.01$, $0.36 \pm 0.02$, and $0.41\pm 0.02$ MK,
 for
E(B$-$V) = 0.01, 0.03, 0.05, and 0.07, respectively;
the corresponding
$\chi_\nu^2$ values are
0.90, 0.87, 0.85, and 0.85, for 10 dof.
An example of best-fit blackbody spectrum is shown in Figure 4,
for E(B$-$V)=0.03.

\begin{figure}[]
 \centering
\includegraphics[width=3.3in,angle=90]{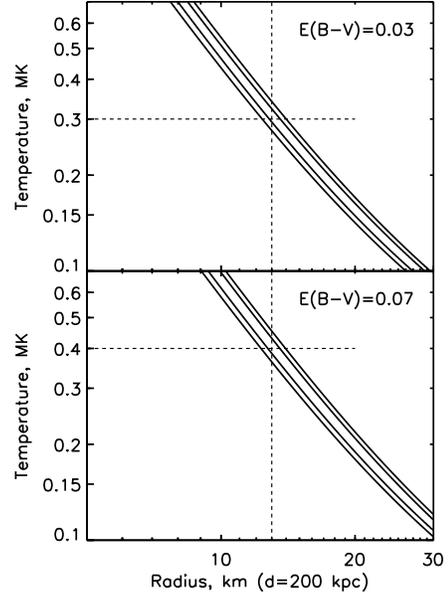}
\caption{ Confidence contours (67\% and 99\%) for the absorbed
blackbody model fit to the FUV-MAMA spectrum, for E(B$-$V)=0.03 and
0.07.}
\end{figure}

\subsection{NIR through FUV spectrum}
To compare the UV emission of Geminga
with its NIR-optical emission,
we plotted in Figure 7 the FUV-MAMA and NUV-MAMA spectral fluxes
$\langle F_\nu \rangle$
together with the fluxes at lower frequencies measured in eight
broad passbands.
Seven of these fluxes have been published previously (see caption to
Fig.\ 7 for references), while the flux marked `555W' in Figure 7
was measured in this work from a recent observation of Geminga with
the {\sl HST} Advanced Camera for Surveys (ACS).

Geminga was observed with the Wide Field Channel (WFC) of the
ACS on 2003 October 7 for 6296 s total exposure (3 {\sl HST} orbits,
two dither positions per orbit, two exposures per dither position)
 in the F555W filter
(ACS `V filter'; pivot wavelength 5358 \AA, FWHM 1235 \AA). We
combined the aligned, pipeline calibrated images from the three
orbits into a single image and performed aperture photometry using
the {\it phot} task from the IRAF {\it apphot} package. To extract
the source counts, we used a circular aperture with a radius of
$0\farcs15$ (3 WFC pixels) which provides an optimal S/N$\approx
36$. For this aperture, the encircled energy fraction of
$0.82\pm0.04$ was determined from the empirical PSF measured for six
field stars with apertures varying from 2 to 25 WFC pixels. The
source count rate was corrected for the finite aperture size and
converted to average spectral flux (cf.\ Eq.\ [3]) using the
conversion factor (inverse sensitivity), $1.974\times 10^{-19}$ erg
cm$^{-2}$ \AA$^{-1}$ per count, for this observing
mode\footnote[6]{see
http://www.stsci.edu/hst/acs/documents/handbooks/DataHandbookv2\\/intro\_ch34.html\#1896082
}. The accuracy of the flux measurement, about 10\%, is limited by
the uncertainty in the encircled energy fraction and various
systematic uncertainties. The flux we measured, $\langle
F_\nu\rangle = 0.17\pm 0.02\,\mu{\rm Jy}$, is a factor of 1.6 lower
than that measured by Bignami et al.\ (1996) from the {\sl HST}
WFPC2 observation of 1994 September 23 with a similar filter.
 We remeasured the WFPC2/F555W
flux and obtained a value consistent with our ACS result within the
uncertainties.
With the new value for the F555W flux
and the other NIR-optical fluxes, we conclude that the ``cyclotron
feature'' in the Geminga's spectrum (Bignami et al.\ 1996; Mignani et
al.\ 1998) was likely a result of inaccurate photometry.

\begin{figure}
\centering
\includegraphics[width=2.6in,angle=90]{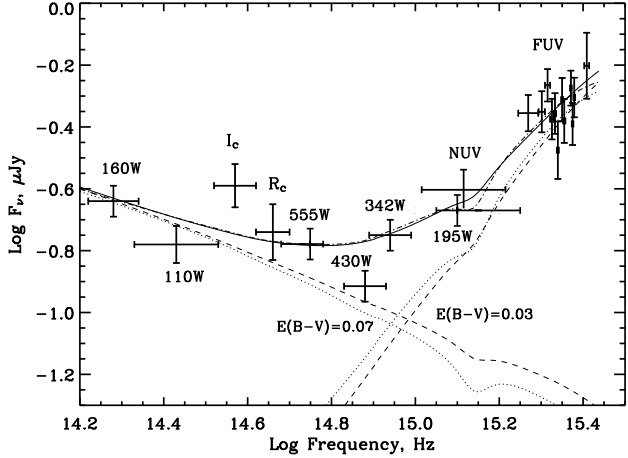}
\caption{NIR through FUV spectrum of Geminga. The broadband fluxes
were measured with the {\sl HST} NICMOS (F110W and F160W; Koptsevich
et al.\ 2001), {\sl Subaru} SuprimeCam (I$_{\rm c}$ and R$_{\rm c}$;
Komarova et al.\ 2003), {\sl HST} ACS/WFC (F555W; this work), and
{\sl HST} FOC (F430W, F342W, and F195W; Bignami et al.\ 1996 and
Mignani et al.\ 1998). The solid and dash-dotted lines show the fits
with the absorbed blackbody + power-law model for E(B$-$V)=0.03 and
0.07, respectively. The model components are shown by the dashed and
dotted lines (see text for details).
 }
\end{figure}

It is obvious from Figure 7 that the NIR through FUV spectrum of Geminga
cannot be described by a simple power-law model. We fit this
spectrum with a two-component, power-law plus blackbody, model.
Since the temperature and the radius-to-distance ratio
 of the blackbody component
are strongly correlated in the Rayleigh-Jeans regime ($T R^2/d^2
\approx {\rm const}$), we have to fix one of these parameters in the
fit. For the fixed $R/d = 13\,{\rm km}/200\,{\rm pc}$, we obtained
$T=0.30 \pm 0.02$ MK, $\alpha_\nu = - 0.46\pm 0.12$, $F_0 = 0.11\pm
0.02\,\mu{\rm Jy}$ for E(B$-$V)=0.03,
and $T=0.41\pm 0.02 $ MK, $\alpha_\nu = -0.41\pm 0.13$, $F_0
= 0.12\pm 0.02\,\mu{\rm Jy}$ for E(B$-$V)=0.07 ($\chi_\nu^2 = 1.5$
for 18 dof for each of the fits),
 where $\alpha_\nu$
and $F_0$ are the parameters of the power-law component:
$F_\nu = F_0\, (\nu/1\times 10^{15}\,{\rm Hz})^{\alpha_\nu}$.
Notice that the parameters of the blackbody component are virtually
the same as obtained from the FUV-MAMA spectrum alone.
The best-fit spectra and their components are shown in Figure 7.
We see that the blackbody emission dominates at $\nu \gtrsim 1\times
10^{15}$ Hz ($\lambda \lesssim 3000$ \AA), while the power-law
(presumably magnetospheric) emission dominates at longer wavelengths.

\subsection{Timing analysis}

For the timing analysis of the NUV-MAMA and FUV-MAMA data,
we used the so-called TIME-TAG data files
that contain the photon arrival times, recorded at a 125 $\mu$s
time resolution, and high-resolution detector coordinates (see
\S2.2) associated with each of the events.
We use 2688 NUV-MAMA events extracted from an aperture of
8 high-resolution pixels radius
(includes 66\% of source counts) and 1939 FUV-MAMA events
extracted from the above-defined $\lambda$-bins
with heights of extraction box varying from 14 to
22 high-resolution pixels,
depending on the $\lambda$-bin (includes 46\% of source counts).
The arrival times are corrected for the Earth and spacecraft motions and
transformed to barycentric dynamical times (TDB) at the solar
system barycenter, using the STSDAS task {\it odelaytime}.
The time spans of the FUV-MAMA and NUV-MAMA observations are
19,181 s and 19,785 s, respectively, with a gap of 84,041 s
between the last FUV-MAMA event and the first NUV-MAMA event.

The expected frequency of Geminga's pulsations at the epoch of our observation,
around 52,332 MJD, can be estimated from the previous
timing observations in $\gamma$-rays and X-rays. The most recent
ephemerides of Geminga were published by Jackson et al.\ (2002; J02 hereafter).
These authors found a small glitch in the Geminga's timing history and
presented a post-glitch ephemeris for a time interval of 50,382--51,673 MJD.
Although our observation was taken 659 days after the end of that
interval, extrapolation of this ephemeris to 52,332 MJD predicts
the frequency,
$f_{\rm J02}=4,217,608.6953\, \mu{\rm Hz}$,
with a formal uncertainty of $\pm 0.0013\,\mu{\rm Hz}$,
that is about three orders of magnitude smaller than
 we can achieve in our relatively short observation
(see below).
Therefore, we adopt $f_{\rm J02}$ as an estimate of expected frequency
and look for pulsations in its vicinity, in a
$f_{\rm J02}\pm (2T_{\rm span})^{-1}$ frequency range.

\begin{figure}[]
 \centering
 \vspace{-0.0cm}
\includegraphics[width=2.7in,angle=0]{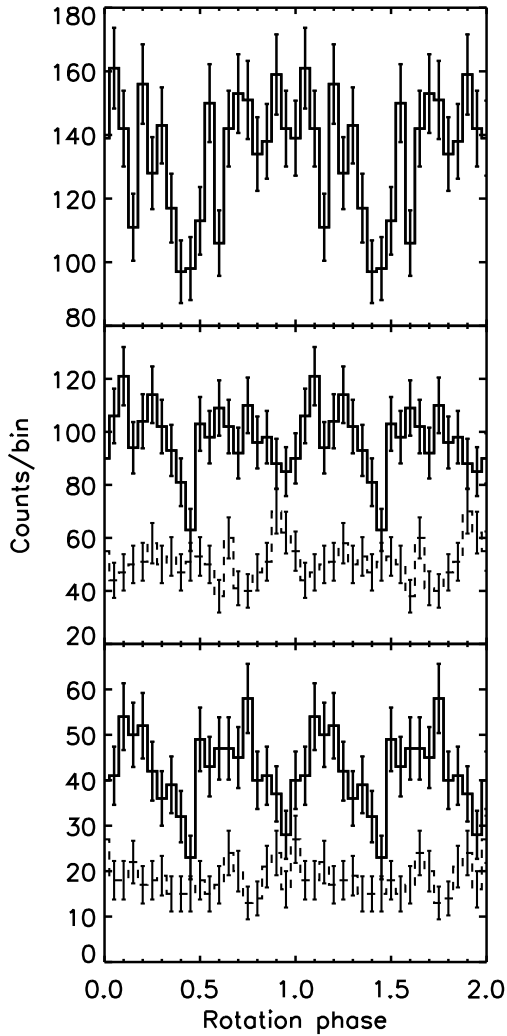}
\caption{UV light curves of Geminga folded with the J02 ephemeris.
  {\em Top}: NUV-MAMA light curve,
obtained using the data from all four orbits. The estimated average
background level in the 20-bin light curve is 44.3 counts per phase
bin. {\em Middle}: FUV-MAMA light curve, obtained using the data
from all four orbits. {\em Bottom}: FUV-MAMA light curve, obtained
using only the data from the first two orbits.  The dashed lines
with associated error bars show the corresponding background light
curves (see text). }
\end{figure}

Since the longer time span, $T_{\rm span} = 123,005$ s,
of the joint  FUV+NUV data set allows a tighter constraint
on the pulsation frequency,
we start from the analysis of this data set.
First, we apply the $Z_n^2$ test (Buccheri et al.\ 1983),
calculating the $Z_n^2$ statistic as a function of trial frequency
in the range of
$f_{\rm J02}\pm 4\,\mu{\rm Hz}$ for $n=1$ through 8, where $n$
is the number of harmonics included. For each of the $n$ values
examined, we
found statistically significant pulsations,
with frequencies of
$Z_n^2$ maxima within ($-$0.3,+0.7)\,$\mu$Hz around $f_{\rm J02}$.
The most significant result
 is obtained
for $n=6$: maximum $Z_6^2 = 53.1$
at $f=4,217,608.8\,\mu{\rm Hz}$; the probability to obtain this value
by chance is $4\times 10^{-7}$.
To better estimate the uncertainty of pulsation frequency,
we also applied the odds-ratio method of Gregory \& Loredo (1992;
see also Zavlin, Pavlov, \& Sanwal 2004) and found
$f=4,217,608.3\,(-0.9,+0.7)\,\mu{\rm Hz}$ for the median frequency
and 68\% uncertainties, and
$f=4,217,608.8\pm 1.2\,\mu{\rm Hz}$ for the mean frequency
and standard deviation.
Within the uncertainties, these frequencies virtually
coincide with the frequency
predicted by the J02 ephemeris.
The expected frequency shift during the FUV+NUV observation,
$\dot{f} T_{\rm span} = 0.024\,\mu{\rm Hz}$, is much smaller
than the frequency uncertainties in our measurement, which
means that this observation is not sensitive to frequency
derivative $\dot{f}$.

Since our spectral analysis has shown that
the FUV radiation is predominantly thermal, while the NUV radiation
has a significant contribution from the magnetospheric component
(see \S2.3 and Fig.\ 7), one can expect different strength and shape of
 pulsations
in the FUV and NUV bands.
Therefore, we analyzed these two data sets separately.
Since the frequency we measured from the FUV+NUV
data is consistent with the J02 ephemeris,
and an {\sl XMM-Newton} observation
taken 37 days later
also suggests that the
ephemeris may still be valid
(see \S3.2), we folded the times of arrival with the J02 ephemeris,
choosing the same zero-phase epoch, 50,382.999999364 MJD.

The folded (source plus background)
 light curve in the NUV-MAMA band, plotted in the upper panel of Figure 8,
shows one broad (FWHM $\approx 0.8$ in phase), flat-top peak per
period, centered at $\phi\approx 1.0$.
The most notable
feature of the pulse profile is the narrow dip at $\phi\approx 0.45$.
The $Z_n^2$ test shows that
the pulsations are statistically quite significant,
with the main contribution coming from the fundamental frequency:
$Z_1^2 = 22.3$ corresponds to $1\times 10^{-5}$ probability
of false result.
The pulsed fraction, defined
as the ratio of the number of counts above the minimum level
to the total number of counts in the light curve, is about 28\%,
which corresponds to the intrinsic source pulsed fraction $f_{\rm p}\approx
40\%$.

The $Z_n^2$ test for the FUV-MAMA data set
shows most significant pulsations for $n=4$:
$Z_4^2=25.08$ corresponds to 99.84\% ($3.2\,\sigma$) significance.
The lower significance of the FUV pulsations, compared to the NUV
pulsations,
can be caused by a lower S/N
in the spectroscopic mode.
Since the thermal-glow background was growing with increasing the
LVPS temperature in the course of our observation (see \S2.2), we
performed the timing analysis for various combinations of orbits
and found that indeed the pulsations were more significant in
earlier orbits. For instance, in the first two orbits ($T_{\rm
span}=
7638$ s,
$N=844$ counts)
the most significant $Z_2^2=23.46$ corresponds to
99.990\% ($3.9\sigma$) significance.

The four-orbit and two-orbit FUV-MAMA light curves
 are shown in the middle and lower panels of Figure 8.
In the same panels we show the light
curves for the background counts extracted from two
boxes centered at $Y=180$ and $Y=240$ high-resolution pixels,
with the same heights as used for extraction of the source events.
The background light curves do not show statistically significant
pulsations.
Both the four-orbit and two-orbit
source-plus-background light curves show a sharp, asymmetric dip
at approximately the same phase as the NUV-MAMA
light curve.
A hint of a shallower dip, better pronounced in the two-orbit light curve,
is seen at $\phi\approx 0.95$.
The pulsed fraction in the observed (source + background)
radiation is about 35\% and 45\%
 for the four orbits and first
two orbits, respectively.
The corresponding intrinsic pulsed
fractions are rather high, about 60\%--70\%.
It should be noted, however, that these values are rather uncertain
because of the large statistical error of the minimum level.

\section{X-ray spectrum and pulsations of Geminga}
To better understand the UV spectrum and pulsations of Geminga,
observations at X-ray wavelengths are particularly useful. The
deepest observation of Geminga in X-rays was carried out with the
{\sl XMM-Newton} observatory on 2002 April 4--5 (orbit 425). The
EPIC\footnote[7]{European Photon Imaging Camera.}-MOS and EPIC-pn
instruments observed the pulsar for 101.4 and 71.4 ks of effective
exposures, respectively. Two EPIC-MOS detectors were operated with
medium filters in Full Frame mode providing an image of a large
area, $r\sim 14'$, with time resolution of 2.6 s. EPIC-pn was used
in combination with thin filter in Small Window mode which covers a
$4\farcm37\times 4\farcm37$ region and provides a $5.7$ ms time
resolution. First results of this observation have been reported by
Zavlin \& Pavlov (2004a) and Caraveo et al.\ (2004a,b). Here we
briefly describe the X-ray spectrum and pulsations of Geminga, with
emphasis on the properties most useful for the comparison with the
optical-UV data.

\subsection{
X-ray spectrum}
The most detailed X-ray spectrum of Geminga was obtained with the
EPIC-pn instrument. The EPIC-pn data, processed with the SAS
package\footnote[8]{{\tt http://xmm.vilspa.esa.es}} (ver.\,6.0.0),
were used for the spectral and timing analysis. We extracted the
source (plus background) photons from a $40''$-radius circle
centered at the pulsar position, which contains about 88\% of source
counts. The estimated total source count rate (corrected for finite
extraction radius) is $0.813\pm0.004$ counts s$^{-1}$ in the 0.2--10
keV range for single and double events (with photon-induced charge
detected in a single CCD pixel and two adjacent pixels). The 0.2--10
keV phase-integrated spectrum was binned in 222 spectral bins with
at least 40 source counts per bin. The detector response matrix and
effective area were generated with the {\em rmfgen} and {\em arfgen}
tools, respectively. Fitting this spectrum with a two-component,
blackbody (BB) + power-law (PL), model, we find the blackbody
temperature $T_{\rm bb}= 0.47\pm 0.02$ MK and radius $R= (17.0\pm
2.5)\,d_{200}$ km, which suggests that the thermal component
originates from the NS surface. The PL component, with a photon
index $\Gamma = 2.02\pm0.05$ dominates at energies $E\gtrsim 0.6$
keV and contains about 10\% of the total luminosity in the 0.2--10
keV band, $L_{\rm 0.2-10\,keV}^{\rm pl}= (2.6\pm0.1)\times 10^{30}
d_{200}^2\,{\rm ergs}\,{\rm s}^{-1} \simeq 7\times 10^{-5} d_{200}^2
\dot{E}$. Extrapolated into the optical domain, the PL component
exceeds the observed optical fluxes by a factor of 100--500, which
might be interpreted as a flattening of the pulsar magnetospheric
spectrum at lower photon energies. The hydrogen column density
derived from this fit is $n_{\rm H}=(2.9\pm 0.2)\times 10^{20}$
cm$^{-2}$.

\begin{figure}[]
\centering
\includegraphics[width=2.7in,angle=-90]{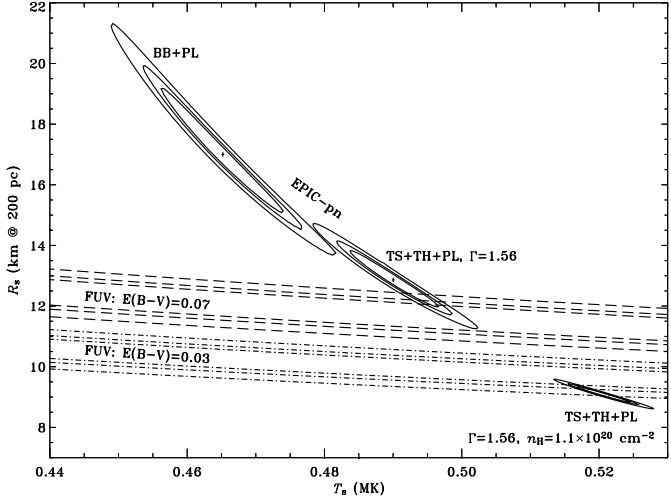}
\caption{Confidence contours (68\%, 90\%, and 99\%)
 in the temperature-radius plane obtained
from fitting the EPIC-pn spectra (solid lines) with the TS+TH+PL and
BB+PL models (labels near the contours). The TS+TH+PL contours were
obtained at the fixed parameters of the PL component; $n_{\rm H}$
was free for the upper contours, while it was fixed at the value
obtained from {\sl ROSAT} data for the lower contours. All the model
parameters were free for the EPIC-pn BB+PL contours. The dashed and
dash-dotted lines show the confidence contours obtained from fitting
the FUV-MAMA spectrum with a blackbody model for two values of the
color index E(B$-$V).
 }
\end{figure}

Although the two-component model cannot be rejected based on the
overall fit quality ($\chi^2_\nu = 1.11$ for 217 dof;
systematic errors in the EPIC-pn response not included),
the fit residuals show
some excess of observed counts over the best-fit model at higher energies,
$E\gtrsim 7$ keV,
indicating a harder PL spectrum. Indeed, fitting the high-energy tail
($E>2.5$ keV)
of the spectrum with a single PL model
gives
$\Gamma = 1.56\pm 0.24$,
$L_{\rm 0.2-10\, keV}^{\rm pl} = (2.2 \pm 0.2) \times 10^{30}d_{200}^2$ erg s$^{-1}$
($\chi^2_\nu = 0.98$ for 30 dof).
The BB+PL fit
with $\Gamma$ fixed at this value
is statistically unacceptable
($\chi_\nu = 2.95$ for 219 dof).
Therefore, we tried a three-component model
consisting of
soft (TS) and hard (TH) blackbody components
and a PL component.
With the
PL parameters fixed at the
values obtained from the best PL fit in the 2.5--10 keV band
($\Gamma = 1.56$, ${\cal N} = 5.5\times 10^{-5}$
photons cm$^{-2}$ s$^{-1}$ kev$^{-1}$ at 1 keV),
we obtain the following parameters for the
thermal components:
$T_s = 0.49\pm 0.01$ MK, $R_s=(12.9 \pm 1.0)\, d_{200}$ km,
$T_h = 2.32\pm0.08 $ MK, $R_h = (46 \pm 12)\, d_{200}$ m, and
$n_{\rm H} = (2.4\pm0.2)\times 10^{20}$ cm$^{-2}$
($\chi_\nu^2 = 1.10$ for 217 dof).
Fitting the EPIC-MOS spectra obtained in this observations yields
almost the same model parameters (discarding the EPIC-MOS events
below 0.3 keV, where the responses of the MOS detectors are known
very poorly).

In the TS+TH+PL model the TS component can be interpreted as emission
from the bulk of NS surface,
the TH component can be ascribed to
emission from smaller, hotter regions of the NS surface,
and the PL component represents the magnetospheric radiation.
Such an interpretation of the Geminga's X-ray spectrum
is in line with the results
obtained from  {\sl Chandra\/} and {\sl XMM-Newton\/}
observations of the other bright middle-aged pulsars,
B0656+14
and B1055--52, whose X-ray spectra can also be desrcibed by the
TS+TH+PL model with similar parameters
(Pavlov et al. 2002;
Zavlin \& Pavlov 2004a).
However, the effective radius of the
Geminga's TH component is much smaller than those of B0656+14
and B1055--52, about $0.6\,(d/0.3\,{\rm kpc})\,{\rm km}$ and
$0.4\,(d/0.7\,{\rm kpc})\,{\rm km}$, respectively.

The confidence contours for the temperature and radius of the TS component
are shown in Figure 9. In the same figure we plotted the
temperature-radius confidence contours obtained from the blackbody
fit of the FUV-MAMA spectrum, for E(B$-$V)=0.03 and 0.07
(see \S2.2 and Fig.\ 6).
We see that at plausible values of interstellar extinction,
E(B$-$V)$\lesssim 0.07$,
 the FUV contours lie at smaller radii
(or lower temperatures) than the X-ray contours. This means that the
extrapolation of the thermal X-ray component into the UV-optical
goes above the observed FUV flux: $(TR^2)_X/(TR^2)_{FUV} \simeq
1.8$, 1.6, 1.4, and 1.2 for E(B$-$V) = 0.01, 0.03, 0.05, and 0.07,
respectively, for the best-fit parameter values (see also Fig.\ 10).
Such behavior is in contrast to other neutron stars observed in both
X-rays and optical, for which such an extrapolation usually {\em
underpredicts} the optical-UV fluxes\footnote[9]{A possible
exception is the Vela pulsar (see Romani, Kargaltsev, \& Pavlov
2005).} by a factor of 2--7 (see \S4.1). To obtain a similar ratio
for Geminga, one would have to assume unrealistically high
extinction, E(B$-$V) $\sim 0.2$--0.4.

\begin{figure}[]
\centering
\includegraphics[width=3.3in,angle=180]{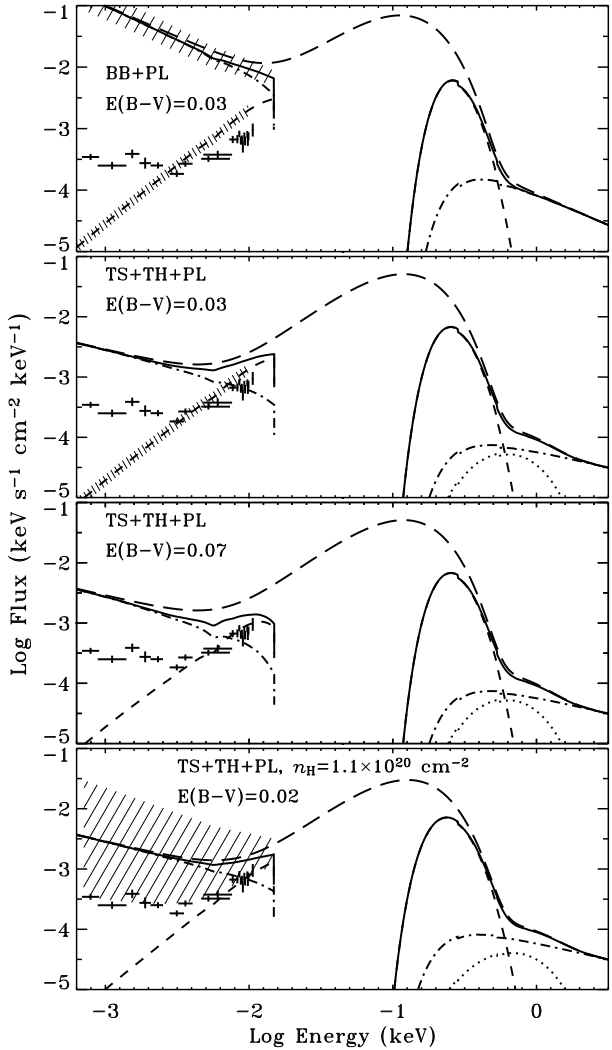}
 \caption{NIR through X-ray spectrum of Geminga for
different X-ray spectral models and different color indices. The
solid lines show the best-fit (absorbed) spectra in the X-ray range
and their extrapolations into the NIR-FUV range. The short-dash and
dash-dot lines show the (soft) thermal and PL components,
respectively, the dotted lines in three lower panels show the TH
component (its contribution is negligible in the NIR-FUV range), and
the long-dash lines present the unabsorbed total spectra. The
crosses depict the measured NIR-FUV spectral fluxes (cf.\ Fig.\ 7).
The hatched areas along the PL and thermal components in the NIR-FUV
range demonstrate propagated uncertainties of the corresponding
extrapolations. The upper panel shows a two-component (BB+PL) X-ray
fit, while three lower panels show TS+TH+PL fits with fixed
parameters of the PL component. The fit shown in the lower panel was
obtained at a fixed $n_{\rm H}$ value, while $n_{\rm H}$ was a
fitting parameter in three upper panels. (See text for more
details.) }
\end{figure}

If we adopt the above-described two-component (BB+PL) model, the
discrepancy between the X-ray and FUV temperature and radius is even more
pronounced, as demonstrated by the corresponding confidence contours
in the upper left part of Figure 9
 and the  upper panel of Figure 10.
We note, however, that
the $n_{\rm H}$ values corresponding to these contours,
$(2.4\pm 0.2)$ and $(2.9\pm 0.2)\times
10^{20}$ cm$^{-2}$ for the TS+TH+PL and BB+PL models,
respectively, significantly exceed the $n_{\rm H} \simeq (1.1\pm 0.2)
\times 10^{20}$ cm$^{-2}$ obtained from the {\sl ROSAT} PSPC observations
(Halpern \& Wang 1997), which indicates a discrepancy between
the PSPC and EPIC
 responses at low energies. If we fix $n_{\rm H}$
at the best-fit PSPC value, the
confidence contours
shift to higher temperatures and lower radii
(see the lower-right EPIC-pn contours in Fig.\ 9), overlapping the FUV
contours. For this $n_{\rm H}$, the FUV fluxes
lie on the extrapolation
of the best-fit X-ray TS component at E(B$-$V)\,$\lesssim 0.04$
 (see lower panel of Fig.\ 10).
Since neither EPIC nor PSPC have been accurately calibrated for very soft
spectra, systematic errors can substantially exceed the statistical
errors, and the model parameters inferred from such fits may not
be very accurate. Therefore, there still remains some uncertainty in
the comparison of the X-ray (Wien) and UV (Rayleigh-Jeans) tails of
the thermal spectrum.
However, even with account for this uncertainty, Geminga exhibits
a fainter UV-optical thermal radiation, relative to the
soft X-ray radiation, than the other neutron stars
for which such a comparison is possible (see \S4.1).

Figure 10 also shows that the continuation of the
best-fit X-ray PL into the
optical very strongly overpredicts the observed NIR-optical fluxes
for the BB+PL model. However, the predicted and observed fluxes
become marginally consistent if we use the PL component
inferred from the $E >2.5$ keV spectral tail.
\subsection{X-ray pulsations}
To study the X-ray pulsations of Geminga, we use
the same EPIC-pn data
($T_{\rm span}=101.9$ ks, epoch of the middle of the time span
52,369.2997 MJD).
First, we measured the pulsation frequency using
the $Z_n^2$ and odds-ratio methods,
 for various energy bands and
extraction radii, and found
most probable frequencies in the range of
4,217,607.75--4,217,607.96 $\mu$Hz,
with typical uncertainties of about $0.1\,\mu{\rm Hz}$
 for individual measurements.
For example, the odds-ratio method applied for 42,170
events in the 0.23--4.0 keV band, extracted
from $40''$-radius circle, gives
$f=4,217,607.85\pm 0.10\,\mu{\rm Hz}$ for the mean frequency
and standard deviation, and
$f=4,217,607.86\, (-0.05,+0.05; -0.16,+0.12; -0.24,+0.19)\,\mu{\rm Hz}$
for the median frequency and 68\%, 90\%, and 99\% uncertainties.
The most probable frequencies are consistently lower,
by 0.1--0.3 $\mu$Hz,
than the frequency
$f_{\rm J02}=4,217,608.0664\pm 0.0013\,\mu{\rm Hz}$ predicted by the J02
ephemeris.
However, since the differences do not exceed 3\,$\sigma$ uncertainties
of our measurements, it is still possible that the J02 ephemeris is
applicable at the epoch of the {\sl XMM-Newton} observation.

\begin{figure}[]
\centering
\includegraphics[width=3.3in,angle=180]{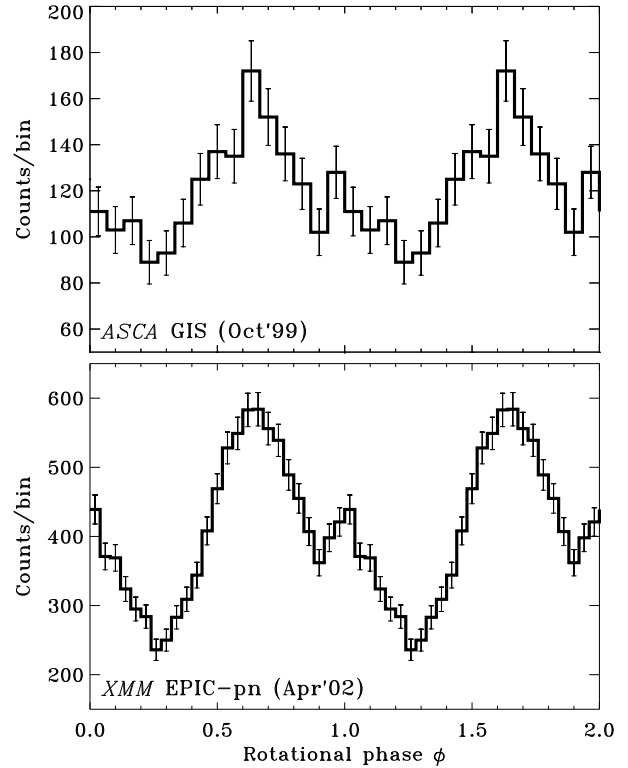}
 \caption{{\sl ASCA} GIS and {\sl XMM} EPIC-pn light curves in the
0.5--2 keV range folded with the J02 ephemeris. }
\end{figure}

We have also directly checked the phase alignment of the light curves
extracted from the {\sl XMM-Newton} data with those observed by {\sl ASCA}
in 1999 October 5--11.
The lower panel of Figure 11 shows the {\sl XMM-Newton}
 light curve folded with the
J02 ephemeris, in the energy band 0.5--2 keV
(10,264 counts in $40''$-radius aperture). The upper panel
of the same figure shows the light curve obtained  with
the two {\sl ASCA} GIS instruments
($T_{\rm exp}=207.8$ ks, $T_{\rm span}=486.5$ ks,
epoch of the middle of the time span MJD 51,459.7356;
1,819 counts in $3'$-radius
aperture)
folded with the same ephemeris and in the same energy range.
We see that not only the shapes of these light curves are virtually
the same, but also their phases are in excellent agreement,
within the phase uncertainty ($\simeq 0.12$) of the J02 timing solution
propagated to the epoch of the {\sl XMM-Newton} observation.
Therefore,
we will assume that the J02 ephemeris
is still applicable in 2002 April
and use it to compare the light curves observed with
different instruments.

\begin{figure}[]
\centering
\includegraphics[width=3.3in,angle=0]{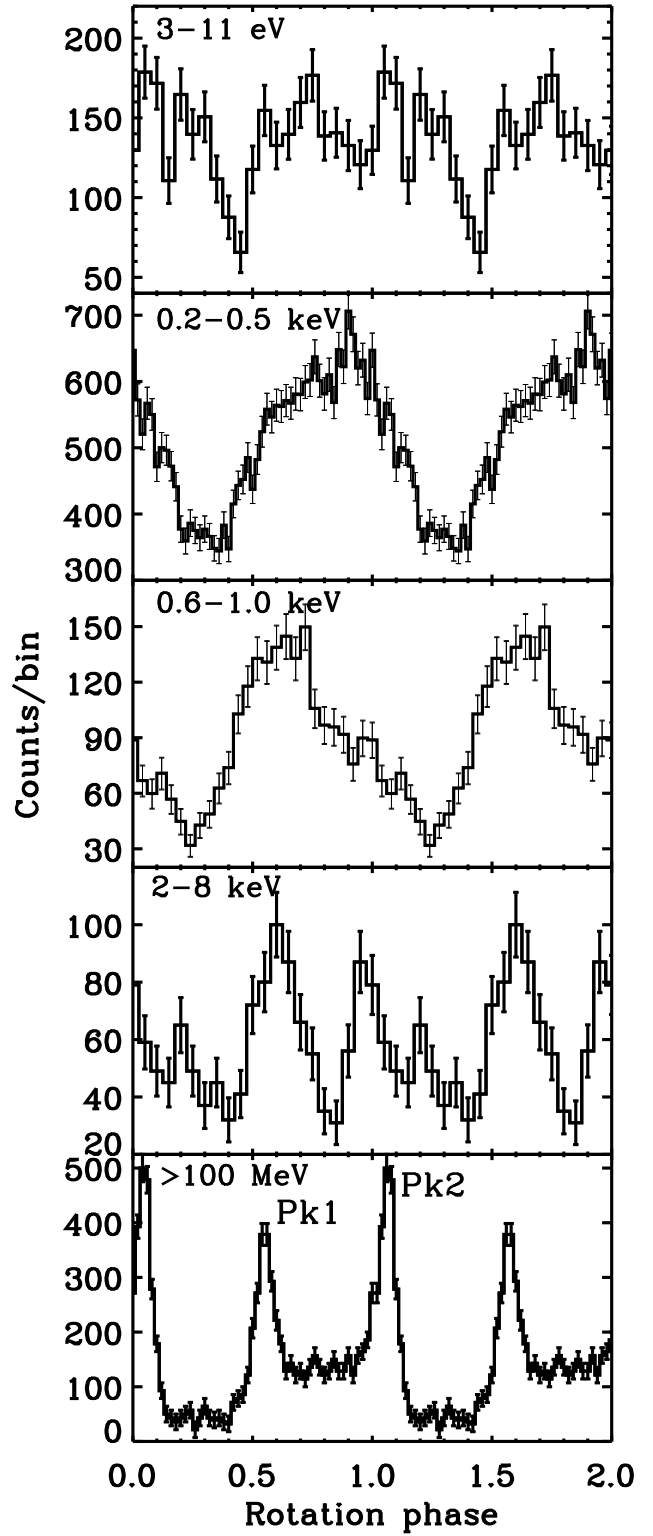}
\caption{Background-subtracted light curves of Geminga
 in UV (NUV-MAMA + FUV-MAMA),
X-rays (EPIC-pn) and $\gamma$-rays (EGRET) bands, folded with the
J02 ephemeris. The $\gamma$-ray light curve is taken from Jackson et
al.\ (2002). }
\end{figure}

The background-subtracted light curves in the energy ranges
0.2--0.5, 0.6--1.0, and 2--8 keV are shown in Figure 12. In the
0.2--0.5 keV and 2--8 kev bands the radiation is dominated by the TS
and PL components, respectively, while the 0.6--1.0 keV band was
chosen around the maximum of the TH component (see Fig.\ 10). The
light curves were extracted from a smaller $30''$-radius aperture
and a shorter, 80.0 ks, time span (exluding intervals of strong
background flares at the beginning and end of the observation) to
reduce the background contamination and maximize the signal-to-noise
ratio. The 2--8 keV light curve (pulsed fraction $f_{\rm p}=34\%\pm
8\%$) shows two pronounced peaks per period, resembling the
$\gamma$-ray light curve (albeit with smaller distance between the
peaks) and a hint of a third peak, at $\phi \approx 0.2$. On the
contrary, the 0.2--0.5 keV light curve ($f_{\rm p} = 30\% \pm 2\%$)
is characterized by one broad peak per period (with small
``ripples'', perhaps due to contribution from the PL and TH
components). The 0.6--1.0 keV light curve shows the highest pulsed
fraction, $f_{\rm p} = 62\%\pm 5\%$, with one asymmetric peak,
possibly comprised of several peaks associated with contributions
from different components in this band (Fig.\ 10). The minimum of
the 0.2--0.5 keV light curve is approximately aligned in phase with
one of the minima of the 2--8 keV light curve, being shifted by
$\Delta\phi \approx 0.1$ from the sharp dips of the NUV and FUV
light curves. (One should remember, however, that the shift can be
caused by errors in phase alignment.) Examples of X-ray light curves
for other energy ranges can be found in Zavlin \& Pavlov
(2004a)\footnote[10]{ The inaccurate estimates of pulsed fractions
given in Figure 8 of Zavlin \& Pavlov (2004a) should be
disregarded.} and Caraveo et al.\ (2004a,b). A more detailed
discussion of the thermal and nonthermal light curves is presented
in \S4.

\section{Discussion}
The above-described results of the observations of Geminga show
that both the NIR-optical-UV and X-ray emission are comprised of
thermal and nonthermal components, with quite different spectra
and light curves. In the following,
we discuss the multiwavelength properties
of these components separately.

\subsection{Thermal
component(s) of the Geminga's emission}
\subsubsection{Spectrum}
It follows from \S2.2 and \S3.1 that the Geminga's radiation is
predominantly thermal
at $4\,{\rm eV} \lesssim E \lesssim
0.5\,{\rm keV}$.
The observed FUV
and soft X-ray (TS component) spectra represent the
Rayleigh-Jeans and Wien tails of the thermal spectrum
emitted from the NS surface. The
blackbody fits of the soft X-ray emission give the NS surface temperature
in a range of 0.45--0.53 MK ($\approx 39$--46 eV),
corresponding to effective radii of 21--8 km, at $d=200$ pc
(see Figs.\ 9 and 10).
The uncertainty in these parameters is mostly
due to the poorly calibrated responses of the EPIC
detectors at low energies.
Moreover, these
 temperatures are somewhat lower, and the radii larger, than those
estimated from the previous {\sl ROSAT} and {\sl EUVE} observations
(Halpern, Martin, \& Marshall 1996; Halpern \& Wang 1997).
Since significant variations of the NS temperature
and emitting area in a ten year span of these observations
can hardly be expected,
the discrepancy is most likely due to discrepant instrument responses
at low energies.

\begin{figure}
\centering
\includegraphics[width=2.7in,angle=-90]{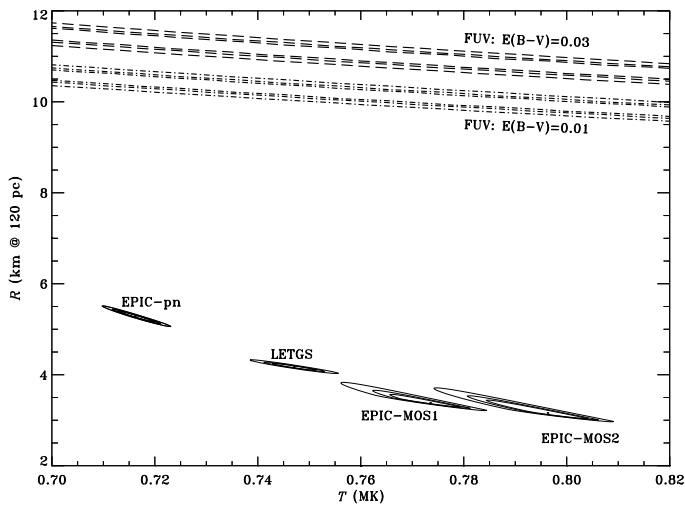}
\caption{Temperature-radius confidence contours (68\%, 90\%, and
99\%) for the isolated neutron star RX J1856.5-3754 obtained from
the X-ray observations with different instruments (solid contours)
and FUV-MAMA observations (dash and dash-dot lines). }
\end{figure}

As we have shown in \S3.1, the observed thermal UV spectrum of Geminga
either matches the continuation of the thermal X-ray spectrum
or lies somewhat below that continuation (up to about one stellar
magnitude), depending on assumed extinction and X-ray spectral model.
On the contrary, other NSs observed in both UV and optical show
relatively brighter Rayleigh-Jeans components, well above the
continuation of the X-ray thermal spectrum. To demonstrate this
difference and show that it is not associated with uncertainties in
instrument responses, we re-analyzed the
 FUV-MAMA and EPIC
data on the best-studied isolated neutron star, RX~J1856.5--3754
(J1856 hereafter; see
Tr\"umper et al.\ 2004
for a recent review of its properties).
We used the FUV-MAMA observation of 2002 October 26 (exposure time 13,451 s),
analyzed the data as described in \S2.2, and confirmed that the
spectrum follows a Rayleigh-Jeans law (Pons et al.\ 2002), with a total flux
$F = (1.89\pm0.09) \times 10^{-14}$ ergs s$^{-1}$ cm$^{-2}$
in the 1155--1702 \AA\ range.
Fitting the FUV-MAMA spectrum
with a blackbody model gives $T=(0.45\pm 0.02)\, R_{13}^{-2} d_{120}^2$ MK
and $(0.55\pm 0.02)\, R_{13}^{-2} d_{120}^2$ MK,
at plausible color indices E(B$-$V)\,=\,0.01 and 0.03
($R_{13} = R/13\,{\rm km}$, $d_{120}=d/120\,{\rm pc}$).
The temperature-radius confidence contours of these fits are shown
in Figure 13.
We also re-analyzed the archival
{\sl XMM-Newton}
observation of 2002 April 8--9
together with the recent observation of 2004 April 17--18.
In the observations of 2002 and 2004, the EPIC-pn was operated in
Small Window mode with thin filter (40.0 ks effective exposure)
and Timing mode with thin filter (64.1 ks effective exposure),
respectively. The EPIC-MOS observations of 2004 were carried out
in Full Frame mode with thin filter (the same 65.3 ks effective exposures
for MOS1 and MOS2). We did not use the EPIC-MOS observations of 2002
because MOS1 was operated in Timing mode, which is very poorly calibrated
for this instrument, and MOS2 was operated in Small Window
mode, with a field-of-view $100''\times 100''$ (for the central CCD)
too small to reliably subtract the background.
We found that the two
EPIC-pn spectra of J1856 are well consistent with each other in
the 0.3--1 keV range (there is no spectral information
available below 0.3 keV in the data collected in Timing mode).
Since J1856 does not show a nonthermal component, we fit the spectra
with a single-component BB model and plot the corresponding confidence
contours in Figure 13 [$n_{\rm H,20}=0.66\pm 0.3$, $0.04\,(-0.04,+0.12)$,
and $0.03 (-0.02,+0.41)$
for the pn, MOS1 and MOS2 detectors, respectively].
In the same figure we also plot the confidence contours obtained
from fitting the 449.9 ks observation of 2001 October 8--15 with the
Low Energy Transmission Grating Spectrometer (LETGS) on {\sl
Chandra} (see Burwitz et al.\ 2003); the corresponding hydrogen
column density is $n_{\rm H,20}=0.86\pm 0.15$. We see that the
observations with different X-ray instruments yield quite different
spectral parameters. This demonstrates once more the lack of proper
cross-calibration of instrument responses to soft spectra and the
fact that systematic uncertainties greatly exceed statistical ones
for spectra with good statistics.

Even with allowance for the uncertainties in instrument responses,
we see from Figure 13 that the UV contours lie well above the X-ray
contours, i.e., the extrapolation of the X-ray blackbody spectrum of
J1856 into the UV-optical range strongly {\em underpredicts} the
observed UV-optical fluxes, contrary to Geminga (cf.\ Fig.\ 9). The
UV-optical excess in the thermal spectra of J1856 and other
so-called X-ray Dim Isolated NSs
(e.g., RX J0720.4$-$3125)
could be explained
assuming that X-rays are emitted from a small hotter area while the
optical-UV radiation is emitted from the bulk of NS surface,
including colder areas invisible in X-rays (e.g.,
Pavlov et al.\ 2002).
Obviously, the apparently smaller UV-emitting area of
Geminga, as compared to the X-ray-emitting area, cannot be explained
by a nonuniform temperature distribution. We might speculate that
the temperature distribution over the bulk of Geminga's surface is
more uniform than in the case of J1856, e.g., because of a different
geometry and strength of the magnetic field that affects the heat
conductivity and, hence, the surface temperature distribution.
However, to explain why the more uniformly heated Geminga exhibits
quite substantial pulsations of its thermal radiation ($f_{\rm
p}\approx 30\%$ in soft X-rays) while no pulsations have been
detected from J1856, one has to  assume a special orientation of
the spin axis of J1856.

Any realistic
interpretation of thermal emission from NSs should take into account
possible deviations of thermal
spectra emitted from NS surface layers
(e.g., atmospheres) from the idealized Planck spectra as well as
the anisotropy of the surface emission associated with strong magnetic
fields (e.g., Pavlov et al.\ 1995; Rajagopal, Romani, \& Miller 1997;
 Zavlin \& Pavlov 2002). For instance, since the X-ray spectrum
emitted from a strongly ionized hydrogen atmosphere is harder than
the Planck spectrum,  a blackbody fit of such a spectrum gives a
temperature exceeding the actual effective temperature by a factor
of 1.5--2.5 and a radius a factor of 3--15 smaller than the actual
radius of the NS, at a given distance. Moreover, the optical part of
the spectrum emitted from such an atmosphere strongly exceeds the
extrapolation of the blackbody fit of its X-ray spectrum into the
optical domain (Pavlov et al.\ 1996). Although the fully ionized
atmosphere models are not applicable to cold NSs (e.g., they give an
improbably large radius-to-distance ratio, $R\sim 100$ km at $d=200$
pc, for Geminga; Meyer, Pavlov, \& M\'esz\'aros 1994), atmospheric
effects might, in principle, explain the large difference between
the observed UV spectrum and the continuation of the blackbody fit
of the X-ray thermal spectrum in J1856 and similar NSs. However, no
realistic models adequately describing the observed broad-band
spectrum of J1856 have been suggested so far, which is not suprising
given the extremely complicated physics of the dense, strongly
magnetized matter at the relatively low temperature of the surface
layers. If we adopt such an interpretation of the strong deviation
of the J1856 broad-band spectrum from a pure blackbody spectrum,
then we have to explain why the broad-band spectrum of Geminga is so
different from that of J1856. Possible hypotheses might involve
different chemical compositions of the surface layers and/or
substantially different magnetic fields (a crude estimate of the
Geminga's magnetic field is $\sim 2\times 10^{12}$ G, but the
magnetic field of J1856 is quite uncertain\footnote[11]{ For
instance, Pavlov \& Zavlin (2003) consider a possibility that J1856
is a millisecond pulsar with a very low magnetic field, $B\sim
10^8$--$10^9$ G, while Tr\"umper et al.\ (2004) suggest that it has
a very strong field $B>10^{13}$ G.}). Moreover, the surface layers
of Geminga and J1856 might be in different phase states. For
instance, one could speculate that the cold surface of Geminga is in
a solid state while the (hotter) surface of J1856 is in a gaseous or
liquid state, which might explain their different spectra. To
distinguish between these possibilities, reliable models for NS
thermal emission at relatively low temperatures, with account for
the contribution of molecules in the opacity of gaseous atmospheres
with strong magnetic fields (Turbiner \& L\'opez Vieyra 2004) and
possible condensation of the surface layers into a liquid or solid
state (van Adelsberg, Lai, \& Potekhin 2004), are to be developed
and compared with the observational data. Until reliable models are
available, the temperatures and radii obtained from applying
simplified models (blackbody, fully ionized atmospheres, partially
ionized atmospheres without molecules) should be considered as crude
estimates only, and any conclusions based on such fits should be
considered with caution. However, although we cannot trust absolute
values of the parameters obtained with the aid of simplified models,
some interesting qualitative results can be obtained from comparison
of the same parameter measured for different NSs. For instance, fits
of the soft X-ray spectra of Geminga and an older pulsar B1055--52
with any model available give a lower temperature for the younger
Geminga, which may have very interesting implications for the NS
cooling models, suggesting different masses of these NSs (Yakovlev
\& Pethick 2004).

As we mentioned in \S3.1, in addition to the thermal soft (TS) component,
the X-ray spectrum of Geminga apparently has a thermal hard (TH) component,
with a much higher temperature, $T_h \approx 2$ MK,
and an apparent
(isotropic)
luminosity
$L_h \sim 4\times 10^{29}$ ergs s$^{-1}$.
Although such a component
has been seen in the spectra of other middle-aged pulsars,
the effective radius, $R_h\sim 50\,d_{200}$ m,
for the Geminga's TH component is surprisingly
small in comparison with the conventional polar cap radius,
$R_{\rm pc} = (2\pi f R^3)^{1/2} c^{-1/2} \approx 300$ m,
suggested by the pulsar models.
Such a small value of $R_h$ might be explained by a projection effect
(if the magnetic axis remains almost perpendicular to the line of
sight in the course of NS rotation),
but this explanation can hardly be reconciled with the high pulsed fraction
at energies where the TH component contribution is maximal
(see \S4.1.2). On the other hand, we should remember that the
TS component was obtained assuming Planck spectra for both thermal
components and a single power law for the magnetospheric spectrum.
Because both these assumptions are not necessarily correct,
we cannot rule out the possibility that the ``TH component'' is simply
associated with
a harder high-energy tail of the surface radiation
(compared to the pure Wien spectrum) or it is due to a steepening of
the slope of the phase-integrated magnetospheric spectrum with
decreasing photon energy (see \S4.2).

\subsubsection{Pulsations in thermal emission}

One of the most intriguing results of our STIS MAMA observations of
Geminga is the strong, non-sinusoidal pulsations in the FUV range, where the
spectrum is dominated by the thermal component, most likely emitted
from the bulk of NS surface.
The shape of the FUV pulsations is different from that of the soft X-ray
pulsations, where the TS component dominates (see Fig.\ 12).
Obviously, neither FUV nor soft X-ray pulsations can
be produced by the locally isotropic blackbody emission.
To explain the unusual pulse shape and the large pulsed fraction of the
thermal FUV and soft X-ray radiation, we have to invoke effects of
strong magnetic field on the angular dependence of NS surface
emission or assume that there is a ``screen'' in the NS
magnetosphere which may partially eclipse the surface emission at
some rotation phases.

In a strong magnetic field, $B\gg 10^{11} (E/1\,{\rm keV})$ G,
when the electron cyclotron energy $E_c$ exceeds the photon energy,
the local emission is essentially anisotropic
(in particular, beamed along the direction of the magnetic field),
which may lead to strong pulsations of the thermal radiation.
The angular distribution and the shape of pulsations
depend on the properties of the emitting region. For instance,
the angular distribution of local emission from a fully ionized
NS atmosphere shows a strong, narrow peak [$\Delta\theta \sim (E/E_c)^{1/2}$]
along the magnetic field (pencil component) and a broad fan-like
component across the magnetic field (Pavlov et al.\ 1994). When
integrated over the visible surface of a NS with a dipole magnetic
field, the angular distribution of NS radiation
is beamed
along the magnetic axis,
even in the case of a uniformly heated NS surface
(Zavlin \& Pavlov 2002).
Such peaks
could explain the soft X-ray pulsations
(at $E\gtrsim kT_{\rm eff}$), including the observed increase
of pulsed fraction with energy. In this hypothesis, the 0.2--0.5 keV
pulse profile (see Fig.\ 12)
can be interpreted as a sum of a smooth thermal component
(the broad thermal peak,
with a maximum at $\phi \approx 0.8$,
corresponds to the closest approach
of the magnetic axis to the line of sight),
and small
``wiggles'' (e.g., at $\phi\approx 0.9$) due to the contribution
of the magnetospheric radiation.
However, at $E\ll kT_{\rm eff}$
the peaks in the model angular distribution
are too low
to be responsible for the observed
FUV pulsations.
On the other hand, as we mentioned above, the fully ionized atmosphere models
are not directly applicable to the cold Geminga, while the partially ionized
atmospheres have not been well investigated.

If the NS surface matter is in a condensed state, we also should expect
an anisotropic emission.
Although the angular distribution of emission from a condensed
surface has not been studied, the examples of spectral emissivity
for several directions, calculated by van Adelsberg et al.\ (2004),
suggest that at least local radiation is beamed along the magnetic field.
To understand whether the radiation from the entire NS surface
can show pulsations similar to those observed from Geminga, the local
specific fluxes should be integrated over the visible NS surface,
for various magnetic field geometries and orientations of the
spin and magnetic axes (see P\'erez-Azor\'in, Miralles, \& Pons 2004
for a few examples).

An alternative explanation for the narrow deep minima in the UV pulse
profiles
could be a partial eclipse by an object co-rotating with
the NS. Since the shapes of the UV and soft X-ray light curves are different
(in particular, the minima are broader in soft X-rays), the eclipsing
object should have a wavelength-dependent effective size.
A natural candidate for such a screen
is the magnetospheric electron-positron plasma which can
absorb the NS surface radiation as a result of
the cyclotron resonance scattering
in a resonance layer,
 where the cyclotron energy is equal to the photon energy
in the rest frame of the electron
 (e.g., Blandford \& Scharlemann 1975).
Two types of models have been discussed for the scattering region:
a stationary nonrelativistic plasma in the closed magnetic field lines zone
(Rajagopal \& Romani 1997; Wang et al.\ 1998; Ruderman 2003)
and streams of ultrarelativistic electron-positron pairs ejected
along the open field lines (e.g., Lyubarskii
\& Petrova 1998, 2000, and references therein).
In the latter case, the effects of the
resonant inverse Compton scattering on the properties of observed
UV/X-ray radiation have not been investigated in detail; however,
crude estimates show that an extremely large pair multiplicity
is needed to reach an optical thickness
of $\gtrsim 1$.
In the case of nonrelativistic plasma
in the closed zone,
which can be supported
against the gravitational force by the thermal radiation pressure
enhanced by the cyclotron resonance (Mitrofanov \& Pavlov 1982;
Rajagopal \& Romani 1997),
 the effects of resonant cyclotron scattering
become significant if the electron/positron number density
is a factor of $\sim 10^2$
 larger than the corotation (Goldreich-Julian) density,
$n_{\rm GJ} \sim 10^{13}$ cm$^{-3}$ for Geminga.
The electron-positron pairs could be supplied from acceleration
zones, but rapid pair production (large multiplicity) is needed to provide
so high densities.
In addition, it remains unclear how the electrons/positrons
would lose the longitudinal momentum to become nonrelativistic
particles (the transverse momentum is essentially nonrelativistic
due to the fast synchrotron/cyclotron losses).
If, nevertheless, there is such a nonrelativistic plasma
screen in
the closed zone,
the wavelength dependence of its optical thickness
depends on spatial distribution of scattering particles.
In particular, the assumption that
the minima in the UV light curves of Geminga are caused by
such a rotating screen implies a significant amount of electron-positron pairs
at a distance of $\sim 15\,R_{\rm NS}$, where the
magnetic field is $\sim 10^9$ G.
One might speculate that the broader minima in the soft X-ray
light curve are also caused by a partial eclipse by the screen.
In this case,
the X-ray resonance layer (at a distance $\sim 5\,R_{\rm NS}$)
should subtend a larger solid
angle
than the UV resonance layer.
Alternatively,
if the soft X-rays pulsations are caused
not by the screen but by the
intrinsic anisotropy of the thermal radition in the strong magnetic field
(see above), then
the magnetosphere is transparent for the X-rays we observe, i.e.
the inner boundary of the plasma screen is located
beyond $\sim 5\,R_{\rm NS}$.

To infer the size of the putative screen and understand the
spatial distribution
of electrons/positrons,
independent
information on the orientation of the spin and magnetic axes
would be very useful. Such information could be obtained from X-ray
pulsations of the TH component emitted by hot polar
caps, which apparently contributes to the X-ray spectrum around
$\sim$0.7 keV
(see Fig.\ 10).
Caraveo et al.\ (2004b) have interpreted the EPIC data
as displaying a varying TH component arising from a rotating
polar cap. However, in our analysis the contribution of this component
to the spectral flux is small and is not clearly seen in the
pulse profile.
Indeed, the
0.6--1.0 keV pulse profile (Fig.\ 12) can be decomposed into
a broad thermal pulse with a maximum at $\phi\approx 0.8$, similar
to the mainly thermal 0.2--0.5 keV pulse, and a narrower nonthermal
pulse centered at the same phase, $\phi\approx 0.6$, as the
higher of the two purely nonthermal
peaks in 2--8 keV light curve. In other words, we see no convincing
evidence for the ``TH component'' (hence, polar caps)
 in the energy-dependent light curves.
More definitive information on the axis orientation
could be obtained from phase-resolved X-ray polarimetry (e.g., Pavlov
\& Zavlin 2000; Lai \& Ho 2003), but it is
not possible with the currently operating X-ray missions.

\subsection{Nonthermal emission}
The multiwavelength observations of Geminga show that its
emission in the NIR-optical,
hard X-rays, and $\gamma$-rays is nonthermal, presumably generated in the
NS magnetosphere.
The comparative analysis of the results of these
observations provides an opportunity
to understand the
mechanisms responsible for the magnetospheric emission
 in different energy bands.

First, we can compare the phase-integrated
X-ray spectra at $E\gtrsim 2$ keV
and optical spectra at $E \lesssim 3$ eV,
where the nonthermal emission dominates.
As shown in \S3.1, the fits to the X-ray spectrum
yield substantially
different slopes of the PL component,
 depending on whether the TH
component is included in the model.
With the data available, we
cannot statistically prove or reject the TH component.
However, we
can extrapolate the PL components of the two different models
(with and
without the TH component)
 to the optical and
compare the extrapolations
with the observed nonthermal spectra
(Figs.\ 7 and 10).
The PL fit of the 2.5--10 keV tail
has about the same slope as the
optical PL component [$\Gamma_X=1.56\pm 0.24$,
$\Gamma_{O} = 1.46\pm 0.12$ and $1.41\pm 0.13$ for E(B$-$V) = 0.03
and 0.07, respectively],
and the extrapolation of the X-ray PL spectrum is
marginally consistent with the optical
fluxes
(the uncertainty of the extrapolation is shown
in Fig.\ 10 [lower panel] and Fig.\ 14 for absorbed
and unabsorbed spectra, respectively).
The NIR through X-ray spectrum can be crudely described by a PL model
with a slope $\Gamma_{OX} \approx 1.3$,
which is smaller than $\Gamma_X$ and $\Gamma_O$, but
the differences are close to 1 $\sigma$ uncertainties of the photon indices.
Similar NIR through X-ray behavior,
with $\Gamma_X\approx\Gamma_O\approx 1.5$, has been seen
in another middle-aged pulsar,
B0656+14 (Pavlov et al.\ 2002; Zavlin \& Pavlov 2004a).
Such behavior implies that
the optical and X-ray emission are generated by the same population
of relativistic particles with a PL energy spectrum,
and by the same (likely synchrotron) mechanism.
This conclusion is supported by the fact that the ratio
of optical-to-X-ray luminosities is about the same for all the
pulsars observed in both X-rays and optical, despite a large
scatter of the X-ray and optical ``efficiencies'', $L_x/\dot{E}$
and $L_O/\dot{E}$ (Zavlin \& Pavlov 2004b).

If the TH component is not included in the fit, the extrapolation
of the PL component ($\Gamma_X = 2.02\pm 0.05$)
of the BB+PL fit of the 0.2--10 keV spectrum
exceeds the optical fluxes by more
than two orders of magnitude (see the upper panel of Fig.\ 10).
This might imply a flattening of the spectrum with decreasing
energy, as observed in the younger Crab and Vela pulsars
(Sollerman et al.\
2000; Romani et al.\ 2005). However, the continuation
toward higher energies
of the NIR-optical PL component crosses the X-ray PL component
at energies $\gtrsim 10$ keV, which
means
a ``double break'' of the spectrum between the optical and X-ray bands
(i.e., $\Gamma$ becomes smaller than $\Gamma_O$ and
then increases to $\Gamma_X$).
Although such behavior has been
suggested (with a considerable uncertainty)
 for the young LMC pulsar B0540--69 (Serafimovich et al.\ 2004),
it was not observed in other pulsars and currently does not look
very plausible.

\begin{figure}[]
\centering
\includegraphics[width=3.4in,angle=180]{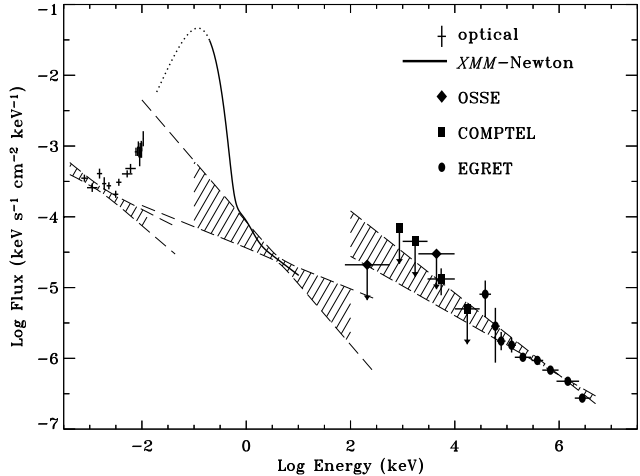}
\caption{Multiwavelength phase-integrated spectrum of Geminga. The
EGRET, COMPTEL, and OSSE points are from Mayer-Hasselwander et al.\
(1994), Kuiper et al.\ (1996), and Strickman et al.\ (1996),
respectively. Fits with the PL model for 3 spectral bands (with $\pm
1 \sigma$ uncertainties) and their extrapolations are shown.}
\end{figure}

One can also extrapolate the best-fit X-ray PL components toward
higher energies to
compare with the $\gamma$-ray data
(Fig.\ 14). The {\sl CGRO} EGRET $\gamma$-ray spectrum has a slope
$\Gamma_\gamma =1.50\pm 0.08$ in the energy range 70 MeV to 2 GeV
(Mayer-Hasselwander et al.\ 1994),
close to that inferred from the PL fit of the 2.5--10 keV tail.
However, the extrapolation of the EGRET spectrum into the X-ray range
exceeds the X-ray PL tail by about one order of magnitude,
while the extrapolation of the X-ray spectrum into
the EGRET range underpredicts the observed flux by a factor of
$\sim$20 at 2 GeV.
(The discrepancy is
of course much larger if we take the PL component of the
BB+PL fit.) Interestingly, the extrapolation of
 the PL that crudely connects the optical
and X-ray points ($\Gamma_{OX} \approx 1.3$) predicts approximately
correct $\gamma$-ray fluxes at $\sim 0.5$ GeV, but the difference in
the slopes is a factor of 2.5 larger than the 1 $\sigma$ error of
$\Gamma_\gamma$. Thus, unless there is a significant systematic
error in the EGRET data analysis\footnote[12]{We should note in this
respect that Grenier, Hermsen, \& Henriksen (1993) reported a
substantially softer spectrum from
 the {\sl COS-B}
observations of Geminga:
$\Gamma_\gamma =
1.84\pm 0.05$ in the 0.05--5 GeV range, or $2.02\pm 0.07$ in 0.14--5 GeV
range, with an indication of a spectral turnover below
$\sim 0.2$ GeV.
With such a soft $\gamma$-ray spectrum, a double
spectral break in between the EPIC and EGRET bands
is certainly required.
On the other hand, Fierro, Michelson, \& Nolan (1998)
found a somewhat harder spectrum
from the EGRET data: $\Gamma_\gamma = 1.42\pm 0.02$ in 0.03--2 Gev
range. Given the large uncertainties in $\Gamma_X$
(including systematic ones: see the
discussion in the next paragraph), the necessity of the double
break becomes less certain.},
 we have to conclude that there should be a double break in the
spectrum between the EPIC and EGRET bands,
which likely means
that  the $\gamma$-ray ray emission
is generated by a different mechanism (e.g., curvature radiation),
or by a different population of relativistic particles, than the
optical/X-ray emission. The {\sl CGRO} COMPTEL and OSSE observations were not
sensitive enough to prove or reject the existence of such a break
(Kuiper et al.\ 1996; Strickman et al.\ 1996).

In the above discussion on the connection between the NIR-optical
and hard X-ray phase-integrated spectra, it was assumed that each
of them can be adequately described by a PL model. In fact,
it is quite plausible that the nonthermal
high-energy radiation of pulsars
is comprised of several
components that are peaked in different directions (phases) and
have different spectral slopes,
as it
has been recently
observed in the hard X-ray emission of the Vela pulsar (Harding et
al.\ 2002) and is routinely seen in the $\gamma$-ray range
(e.g., Fierro et al.\ 1998).
In this case one should expect the phase-integrated
spectrum to have a concave shape (in a log-log scale),
with excesses at low and high energies dominated by
the softer and harder components,
respectively. Some indications of such a concave
spectrum are seen in
the Geminga's nonthermal X-ray component:
while fitting the high-energy tail of the
Geminga's EPIC-pn spectrum with a single PL
and different low-energy cut-offs ($E_{\rm min}$), we noticed that
the best-fit $\Gamma$ increases with decreasing $E_{\rm min}$
(e.g., $\Gamma=1.45\pm 0.30$, $1.56\pm 0.24$, and $1.60\pm 0.20$
 at $E_{\rm min}=3.0$, 2.5, and 2.0 keV, respectively).
Although the change of $\Gamma$ is not statistically significant,
such a trend suggests that the tail is concave
rather than flat, which, in turn, hints that there may be
several emission components (unless the high-energy tail of the
thermal component is harder than in the thermal models
we used). Unfortunately,
the S/N
at $E\gtrsim 2$ keV is too low
to verify this directly with phase-resolved spectroscopy.

It is also interesting to compare the nonthermal pulsations
in different energy bands.
The $\gamma$-ray light curve (lower panel of Fig.\ 12) shows two
peaks per period, at $\phi\simeq 0.55$ (Peak 1) and $\phi\simeq 1.05$
(Peak 2), with a bridge between them.
In the X-ray 2--8 keV band, we also see two peaks, but they are
broader, their separation
($\Delta\phi \approx 0.35$ or $\approx 0.65$)
differs significantly from the $\Delta\phi \simeq 0.5$ in $\gamma$-rays,
and about 60\%-70\% of the 2--8 keV emission is unpulsed, in contrast
to about 100\% pulsed $\gamma$-ray emission.
We cannot determine from these data alone what is the correspondence
(if any) between
the X-ray and $\gamma$-ray peaks
(e.g, the higher X-ray peak could correspond to either Peak 2 or Peak 1,
which would mean that it trails the corresponding $\gamma$-ray peak
by $\delta\phi \approx 0.55$ or 0.05, respectively).
It is tempting to identify the stronger X-ray emission at phases
$\sim$ 0.1--0.5 as a bridge similar to that in the $\gamma$-ray
light curve (which would mean that the higher X-ray peak corresponds to
Peak 2), but the low S/N at these phases
 makes this identification rather uncertain.
Moreover, since no statistically significant change of the peak
phases with energy is seen in the 0.03--3 GeV EGRET range,
it is quite possible that the 2--8 keV X-ray peaks do not directly
correspond to the $\gamma$-ray peaks, being produced by a distinct
mechanism.
Observations at intermediate energies, 10 keV -- 10 MeV,
are needed to clarify
this issue.

As we mentioned above, the highly asymmetric pulse profile at
intermediate X-ray energies 0.6--1 keV can possibly be decomposed into
two components: a broad, likely thermal, component,
centered at $\phi \approx 0.75$, similar
to that seen in the 0.2--0.5 keV band, and a narrower nonthermal peak
around the phase $\phi \approx 0.6$ of the higher
of two 2--8 keV peaks.
However, the putative nonthermal peak at 0.6--1 keV looks significantly
broader than the corresponding peak at 2--8 keV, while the lower
2--8 keV peak is hardly seen in the 0.6--1 keV band.
The 0.2--0.5 keV light curve is clearly dominated by the very broad,
thermal peak centered at $\phi \approx 0.8$;
in the simple dipole geometry of magnetic field
(and the corresponding axisymmetric temperature distribution),
this phase
corresponds to the nearest approach of the only
visible magnetic pole to
the center of stellar disk.
Small contributions from the nontheramal radiation are likely seen
even at these low energies: e.g., some excess emission over
the smooth pulse profile at $\phi\approx 0.55$--0.65 might be due
to the nonthermal peak centered at $\phi=0.6$, while the narrow
``sub-peak'' at $\phi\approx 0.9$ might be associated with the lower
of two 2--8 keV peaks. So, it seems that the two nonthermal X-ray peaks
persist through the whole observed X-ray range, being most clearly
seen in the 2--8 keV band, where they are not ``contaminated'' by the
thermal radiation.

The connection between the optical-UV and X-ray/$\gamma$-ray
pulsations
(see Figs.\ 8 and 12) remains unclear.
We do see two peaks in the FUV band, with about 0.45 phase separation,
but they may not be related to the hard X-ray and/or $\gamma$-ray
pulsations because the FUV spectrum is mostly thermal.
We see only one broad peak in the NUV light curve,
where the nonthermal contribution is
substantial,
$\sim 30\%$--40\%, but we cannot
rule out the possiblity that it is composed of two peaks because
the NUV light curve is very noisy.
Two peaks with a separation of about 0.5 in phase,
similar to the $\gamma$-ray peaks,
were apparently seen in the optical B band
(Shearer et al.\ 1998), but those observations were compromised
by a very high variable background.

It is interesting to compare the multiwavelength
nonthermal radiation of Geminga with theoretical models.
For instance, the outer gap model
by Zhang \& Cheng (2001) predicts one nonthermal X-ray peak
coinciding in phase with the $\gamma$-ray Peak 1
and one soft thermal X-ray peak centered
at the phase of the $\gamma$-ray Peak 2.
However, we see two nonthermal X-ray peaks, while the presumably
thermal X-ray peak is centered in the middle between Peak 1 and Peak 2.
The latter property is expected in the polar cap models for high-energy
radiation, which assume that both nonthermal X-rays
and $\gamma$-rays are emitted along a hollow cone
inscribed in the surface formed by the last open field lines
(Harding \& Muslimov 1998). In this model,
the two nonthermal peaks are seen at the phases when the line
of sight is tangent to the cone at the site where the radiation
is generated (a few NS radii from the surface), which implies
that only one pole is visible (i.e., the angles
between the rotaion and magnetic axes and rotation axis and line
of sight
are $\lesssim 30^\circ$).
The different separation
between the X-ray peaks, in comparison with that of the $\gamma$-ray peaks,
might mean that they are generated at a different
distance from the NS
(e.g., closer to the NS if the higher of the two 2--8 keV
peaks corresponds to the $\gamma$-ray Peak 1)
 while their larger widths suggest that
the X-rays
are less beamed than the $\gamma$-rays.
We might speculate that the two peaks become even broader and closer
in the NUV band, so that they appear as one peak, but
such a trend is not seen in the optical.

Alternatively,
the soft X-ray pulsations may still be magnetospheric in origin.
This is, in fact, natural in the outer magnetosphere picture
of Romani and Yadigaroglu (1995). In their model the closest polar
cap approach is $\delta \phi \sim 0.1$ before the $\gamma$-ray Peak 1,
while the cap producing the observed $\gamma$-ray emission passes
$\delta \phi = 0.5$ later, at $\phi \approx 0.95$ in Figure 12.
These phases are close to the primary and
possible secondary UV minima of Figure 8. In the Dyks and Rudak
(2003) two-pole model, the $\gamma$-ray peaks arise from
opposite hemispheres, but the phase of closest pole approach
is rather similar. In either case, it seems natural to
interpret these minima as the result of scattering screens
above the polar caps removing flux (\S4.1.2). It would be very
surprising if the FUV pulse minimum level represented the true
Rayleigh-Jeans surface flux (with the broad peaks being hot cap
or non-thermal emission), as this would require a remarkably
low NS temperature. In this interpretation there
would be a distribution of soft non-thermal components
dominating the X-ray pulse, mimicking here the TH component
and perturbing the fit of the TS component. Little if any of
the pulsed emission would come from a thermal cap. The
(unpulsed) X-ray surface emission would be about 30\% lower
than that fitted for the TS component, allowing better
agreement with the Rayleigh-Jeans UV flux in Figure 9. Higher
statistics FUV phase-resolved spectra
are required to determine whether the closest approach of the
magnetic axis to the line of sight is associated
with the FUV minima or the soft X-ray maxima.
\section{Summary}
The results of our work
can be summarized as follows.

1.
The STIS/MAMA observations of Geminga
have allowed us to detect, for the first time,
its FUV radiation,
measure the FUV spectrum and NUV flux,
and detect FUV and NUV pulsations.
We also
measured the flux in the ACS/WFC F555W band.
To understand the multiwavelength properties of the Geminga's
radiation, we analyzed its X-ray spectrum and pulsations
observed with {\sl XMM-Newton}.

2.
The phase-integrated
 NIR through FUV spectrum consists of two components,
thermal and nonthermal. The thermal component,
which dominates at $\lambda \lesssim 3000$ \AA,
is emitted from the NS surface.
Its flux corresponds to
a brightness temperature
$T_{\rm RJ} = 0.3$--$0.4\,(d_{200}/R_{13})^2$ MK.
The nonthermal component, which is likely generated in the
pulsar's magnetosphere, shows a PL spectrum with photon
index $\Gamma_O = 1.3$--1.6 and dominates at NIR-optical wavelengths.
We do not confirm the previously claimed spectral feature in the
V band.

3.
The phase-integrated X-ray spectrum is dominated by a thermal
soft component at $E\lesssim 0.5$ keV, with a blackbody temperature
$T_s \approx 0.5$ MK and radius $R_s\approx 13\, d_{200}$ km.
Extrapolation of this spectrum
into the optical-UV domain slightly overpredicts the
thermal component of the observed UV spectrum, in contrast to
all the other NSs observed
in both X-ray and optical-UV ranges. This might be associated
with the very low temperature or a different chemical composition
of the Geminga's surface.
Under the assumption that the blackbody model properly describes
the thermal spectra while the nonthermal spectrum
is a simple power law, an additional thermal component is required
to fit the X-ray spectrum, with a higher temperature, $T_h\approx 2$ MK,
and a small size of the emitting region, $R_h\approx 50$ m.
It is not clear whether this component is real or it
appears because the simplified model spectra were used in the fits.
Indeed,
several soft non-thermal components could contribute to the
X-ray emission, leaving the residual thermal surface flux in
better agreement with the UV Rayleigh-Jeans emission.

4.
The slope of the X-ray nonthermal component, which dominates
at $E\gtrsim 1$ keV, is not well constrained, $\Gamma_X \approx 1.2$--2.0,
because of a large background at these energies and, possibly,
some deviations of the spectrum from a simple PL model.
Its extrapolation into the NIR-optical domain is either marginally
consistent with the observed fluxes or goes above these fluxes,
requiring a flattening of the nonthermal spectrum with decreasing
photon energy. Most likely, the optical radiation and the X-rays
are emitted by the same
population of relativistic particles in
the Geminga's magnetosphere and generated by the same
mechanism.
On the other side, the continuations of
 the nonthermal X-ray and $\gamma$-ray
spectra do not match smoothly, which suggests different radiation
mechanisms in these energy bands.

5.
Unexpectedly, we detected strong pulsations in the FUV band.
The light curve of the
predominantly thermal FUV radiation,
with a narrow, deep minimum and possibly another
minimum shifted by about half a period, differs significantly
from the light curve of the thermal soft X-ray radiation,
which shows one broad, smooth peak per period, possibly distorted
by a small contribution from the magnetospheric component.
Because the FUV pulsations can hardly be explained
by an anisotropic temperature and/or magnetic field distributions,
we suggest that they could be associated with a resonance
scattering of the thermal UV photons in the NS magnetosphere.

6. The light curve of the nonthermal X-ray emission
shows two peaks, as the $\gamma$-ray light curve, but
the X-ray peaks are substantially broader, and their
separation differs significantly from the half-period
separation of the $\gamma$-ray
peaks. At least one of the nonthermal peaks is apparently
seen at lower X-ray energies, superimposed on the
broader peak of thermal emission.
We see no clear connection between the nonthermal
X-ray pulsations and single-peaked NUV pulsations.
Two peaks have been apparently seen in the optical, but
this very noisy light curve is badly in need of confirmation.

To conclude, the {\sl HST}
STIS/MAMA observations
have allowed us to study the UV spectrum and pulsations of
Geminga.
 Combined with the results
of the previous optical and $\gamma$-ray observations and our
analysis of the {\sl XMM-Newton} data, this study has provided a
connection between different energy bands for the thermal and
magnetospheric components and elucidated the multiwavelength picture
of the Geminga's radiation. However, some important properties of
this radiation remain unclear. First, the true nature of the UV
(particularly FUV) pulsations is still uncertain. To understand it,
phase-resolved spectroscopy of the FUV radiation would be
particularly useful, which would require deeper FUV-MAMA
observations, quite feasible with the {\sl HST} STIS if it is
brought back to life in a future servicing mission. Second, it
remains unclear whether we indeed see a small, high-temperature
polar cap or this is an artifact caused by the use of simplified
spectral models. To answer this question we need realistic models
for thermal and nonthermal X-ray radiation and phase-resolved
spectroscopy at $E\gtrsim 0.5$ keV with high S/N. Third, we do not
understand the connection between the nonthermal X-ray emission and
$\gamma$-ray emission. This problem could be resolved by
observations in a 10 keV -- 10 MeV band with a future mission
equipped with detectors more sensitive in this energy range than the
detectors of the {\sl RXTE} and {\sl INTEGRAL} missions. {\sl
NuSTAR}, for example, should be able to offer sensitive measurements
in the hard X-ray range.

\acknowledgements
We thank Rosa Diaz-Miller for the help with the STIS data analysis,
Divas Sanwal for useful discussions, Hans-Albert
Meyer-Haselwander for providing the $\gamma$-ray data,
and Igor Volkov for the help with timing analysis.
The work of V.E.Z. is supported by
a National Research Council Research Associateship Award at NASA MSFC.
Support for programs GO-9182 and GO-9756 was provided by NASA
through a grant from the Space Telescope Science
Institute, which is operated by the Association of
Universities for Research in Astronomy, Inc., under NASA
contract NAS 5-26555.
 This work was also partly supported by NASA grants NAG5-10865,
NAG5-13344, and NNG04GI80G.


\end{document}